\documentclass[10pt,twocolumn,letterpaper]{article}

\usepackage{cvm}
\usepackage{times}
\usepackage{epsfig}
\usepackage{graphicx}
\usepackage{subcaption}
\usepackage{amsmath}
\usepackage{amssymb}
\usepackage{algorithm}
\usepackage{algorithmicx}
\usepackage{algpseudocode}
\usepackage{comment}
\usepackage{amssymb}
\usepackage{soul,color}
\definecolor{Red}{cmyk}{0,1,1,0}
\definecolor{Green}{cmyk}{1,0,1,0}
\definecolor{Cyan}{cmyk}{1,0,0,0}
\definecolor{Purple}{cmyk}{0.45,0.86,0,0}
\definecolor{Rosolic}{cmyk}{0.00,1.00,0.50,0}
\definecolor{Blue}{cmyk}{1.00,1.00,0.00,0}
\definecolor{Orange}{cmyk}{0,0.52,0.80,0}
\definecolor{Black}{cmyk}{1,0,0,1}
\newcommand{\sk}[1]{{\color{Black}            {#1}}} 

\usepackage[pagebackref=true,breaklinks=true,letterpaper=true,colorlinks,bookmarks=false]{hyperref}

\cvmfinalcopy 

\def\cvmPaperID{4} 

\ifcvmfinal\fi
\begin{document}

\title{Geometry-Based Layout Generation with Hyper-Relations AMONG Objects}

\author{Shao-Kui Zhang\\
Tsinghua University\\
{\tt\small zhangsk18@mails.tsinghua.edu.cn}
\and
Wei-Yu Xie\\
Beijing Institute of Technology\\
{\tt\small ervinxie@qq.com} 
\and
Song-Hai Zhang*\\
Tsinghua University\\
{\tt\small shz@tsinghua.edu.cn}
}

\maketitle

\begin{abstract}
    Recent studies show increasing demands and interests in automatically generating layouts, while there is still much room for improving the plausibility and robustness. In this paper, we present a data-driven layout framework without model formulation and loss term optimization. We achieve and organize priors directly based on samples from datasets instead of sampling probabilistic models. Therefore, our method enables expressing and generating mathematically inexpressible relations among three or more objects. Subsequently, a non-learning geometric algorithm attempts arranging objects plausibly considering constraints such as walls, windows, etc. Experiments would show our generated layouts outperform the state-of-art and our framework is competitive to human designers. \footnote{Code is publicly available at \\ https://github.com/Shao-Kui/3DScenePlatform, including the proposed framework (algorithm) and a 3D scene platform (toolbox). }
\end{abstract}

\section{Introduction}
3D scenes are becoming fundamental to many domains in computer graphic, e.g., photo-realistic rendering, virtual reality (VR), providing datasets for computer vision \cite{handa2016understanding}, etc. However, the increasing development of computer graphic requires a better ability to model 3D scenes and provide more layouts. Therefore, we have been investigating techniques of automatically generating scene layouts. 

Generating scene layouts benefits various applications. First, it saves manual work of placing objects, such as video games or industrial designs \footnote{https://planner5d.com/}. Li et al. \cite{li2020automatic} generate various layouts for better simulations of wheelchair training. Handa et al. \cite{handa2016understanding} generate multi-view images from much fewer 3D scenes. 

Existing works already show the progress of scene synthesis \cite{asurveyof3didss}, where scene layouts focus on their plausibility and aesthetic, i.e., visual identifications given generated layouts. Existing works are divided into neural network based techniques and others. The former trains several neural networks for different steps such as placing objects, rotating objects, deciding termination of arrangements \cite{wang2018deep}. The latter formulates a set of mathematical models including graphs, and typically optimize a shuffled area based on e.g., Markov Chain Monte Carlo (MCMC) \cite{yu2011make,qi2018human}, since the models are too complicated to be solved. Nevertheless, algorithmic methods have not been investigated as far as we reviewed, because similarly we have to embed layout rules into an algorithm so that it operates properly. However, layout rules are innumerable. A qualitative comparison of existing techniques is beyond the scope of this paper. Despite underlying technical details, this paper focuses on the final results, i.e., improving the plausibility and aesthetic of generated layouts. 

\begin{figure*}
    \centering
    \includegraphics[width=\linewidth]{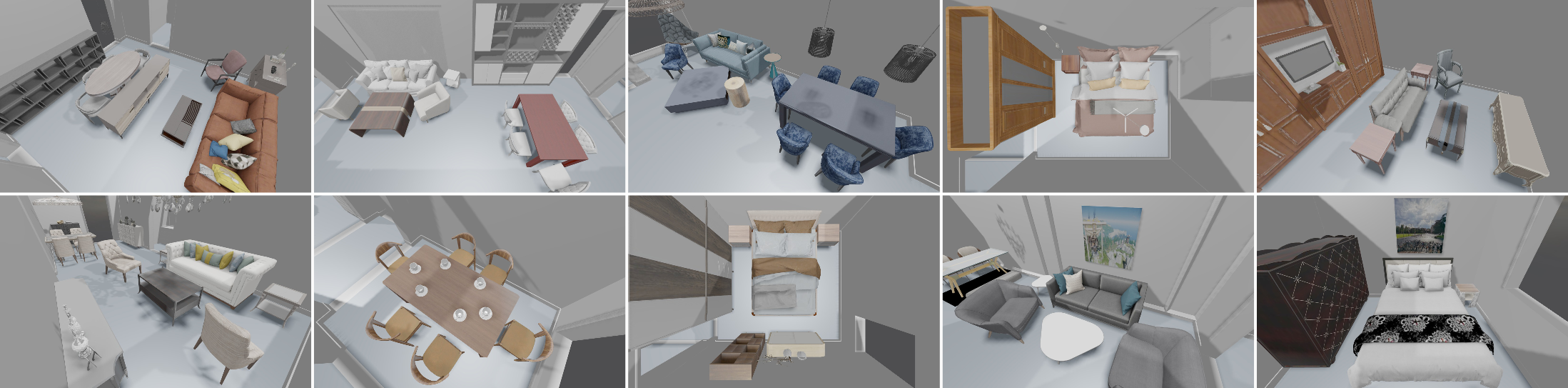}
    \caption{Our framework uniformly layouts objects, e.g., small objects on a surface are arranged concurrently instead of another layout problem. In addition to the overall plausibility, we emphasize reasonableness among objects related to each other, i.e., coherent groups. Ours is also friendly to objects hung on walls. }
    \label{fig:teaser}
\end{figure*}

In this paper, we propose an algorithmic framework for generating room layouts as shown in Fig. \ref{fig:teaser}. Our framework is split into a data-driven phase: coherent grouping and a non-data-driven phase: geometric arranging. In data-driven phase, objects are grouped into several coherent groups (section \ref{sec:def}), where priors are learnt for suggesting layouts within each coherent group. We directly use correct and denoised samples extracted from datasets as priors. This enables two factors. First, we no longer hypothesize distributions of layout rules between/among objects, especially mathematically inexpressible relations. Second, we could easily formulate and represent relations among three or more objects since we only have to structure samples of real distributions. Similar to ``hyper-graphs" where an edge connects to more than two vertices, we name our learnt relations among objects ``hyper-relations" (section \ref{sec:priors}). Thus, several objects of the same coherent group are arranged in $O(1)$ time by sampling their hyper-relations. In non-data-driven phase, given independent coherent groups where objects of the same group are already properly arranged among each other, a geometric algorithm is proposed to generate positions and orientations of each group. Since layout rules among objects are applied during data-driven phase, geometry phase concentrates on much fewer rules related to walls, windows, etc (section \ref{sec:layoutGen}). 

\begin{figure}
	\centering
	\begin{subfigure}[b]{0.32\linewidth}
		\includegraphics[width=\linewidth, height=2.5cm]{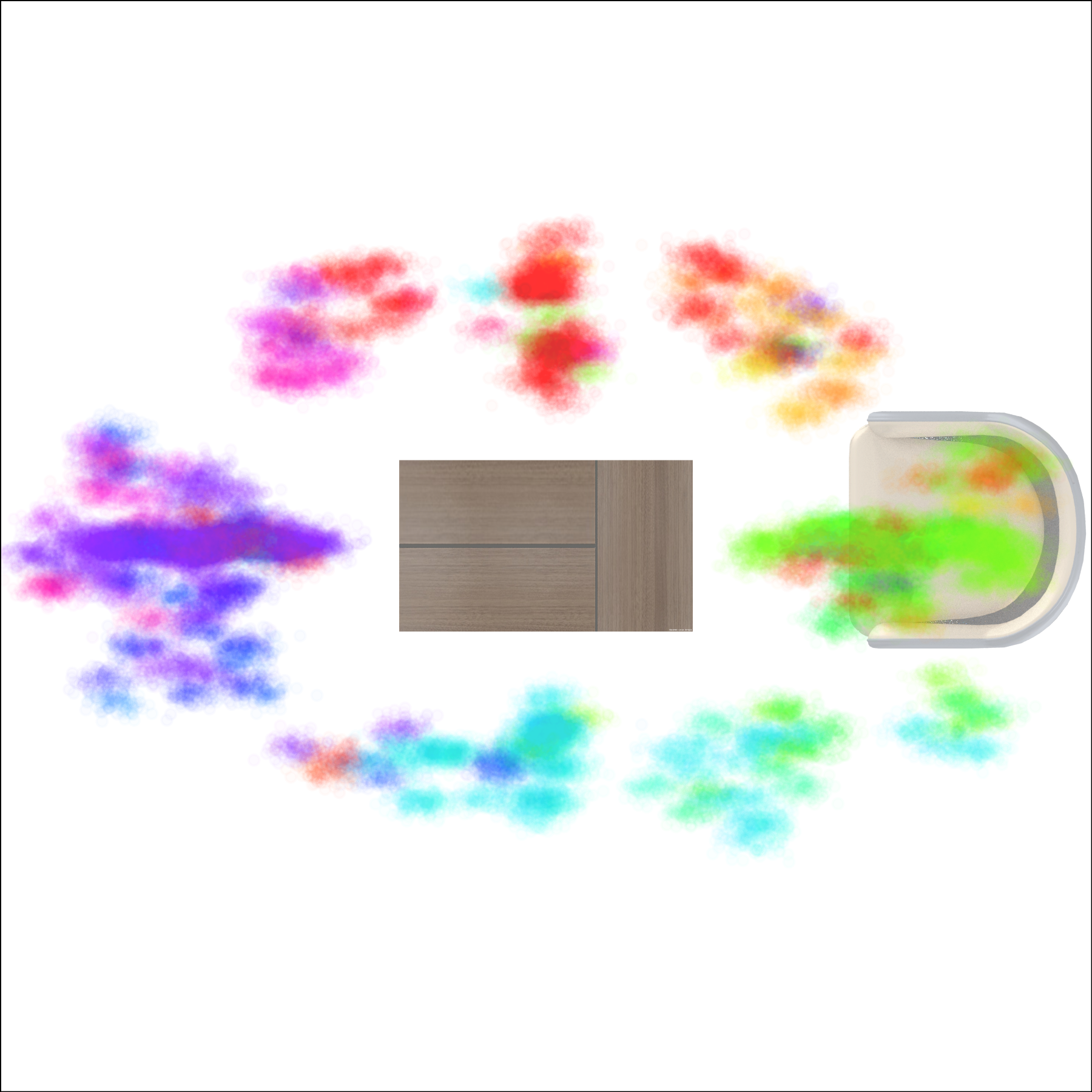}
		\caption{\scriptsize Round Sofa}
	\end{subfigure}
	\begin{subfigure}[b]{0.32\linewidth}
		\includegraphics[width=\linewidth, height=2.5cm]{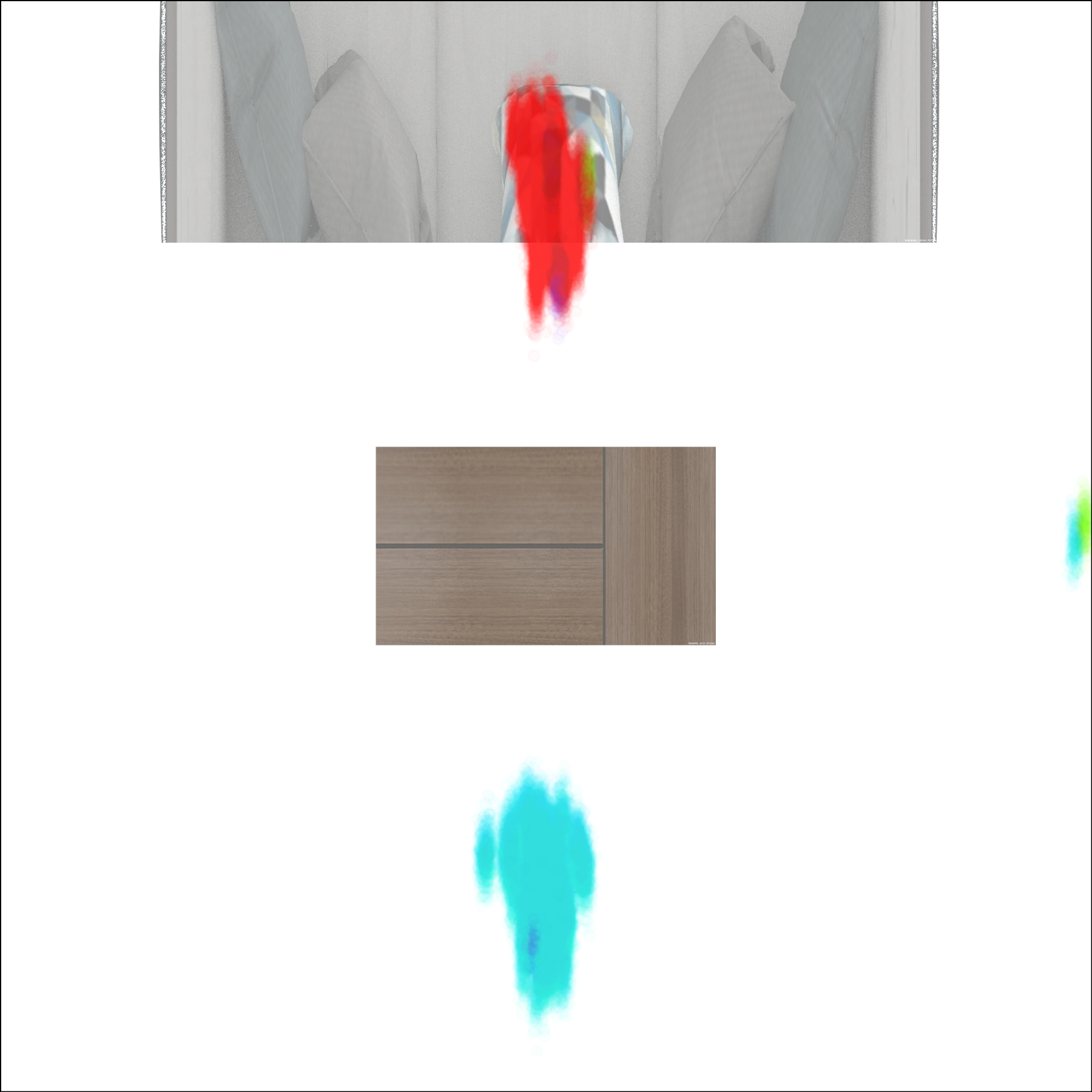}
		\caption{\scriptsize Long Sofa}
	\end{subfigure}
	\begin{subfigure}[b]{0.32\linewidth}
		\includegraphics[width=\linewidth, height=2.5cm]{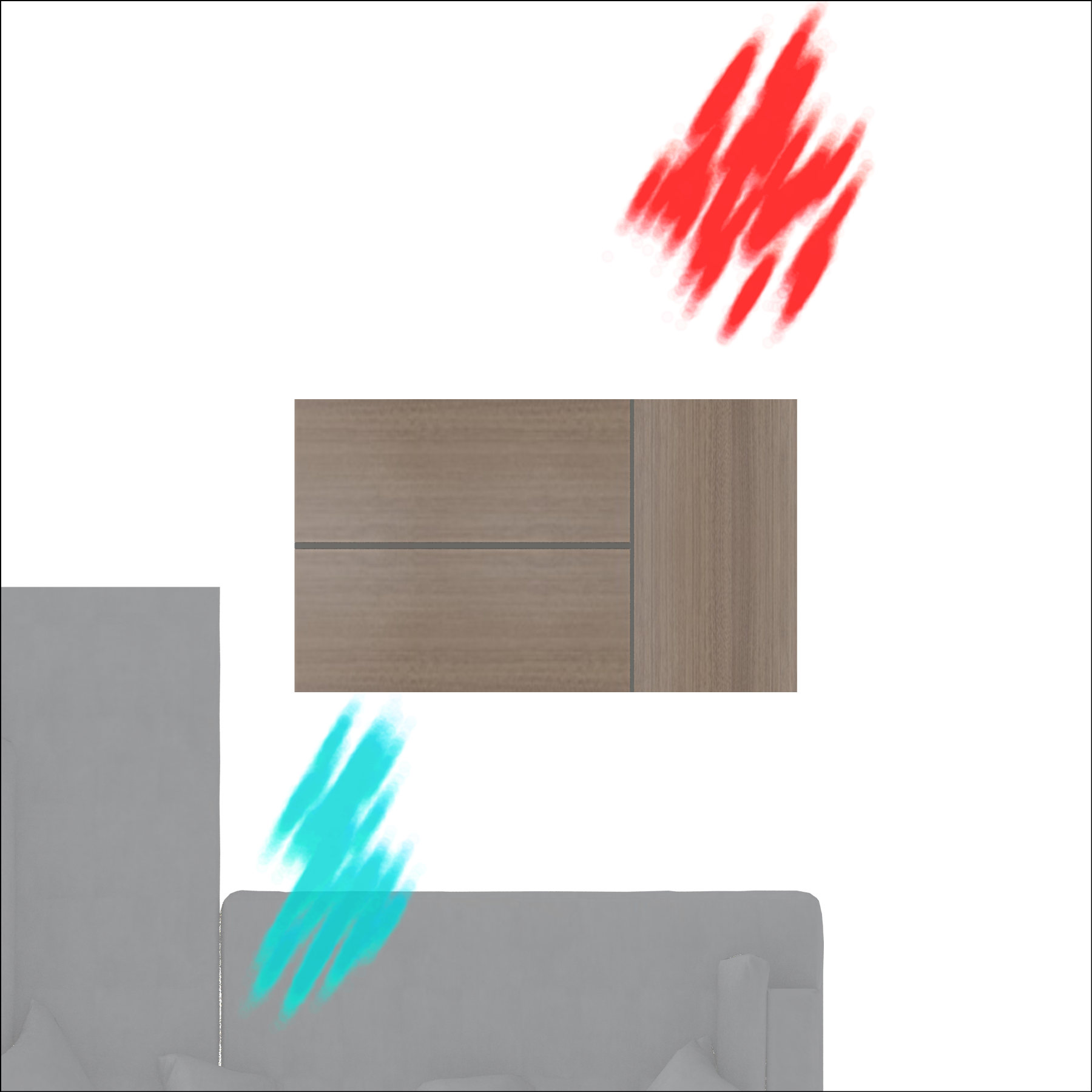}
		\caption{\scriptsize L-Shape Sofa}
	\end{subfigure}
	\caption{Extracting priors based on instances results in finer priors. This figure shows the priors of the same coffee table with respect to three sofa instances of different shapes and geometries. }
	\label{fig:whyInstance}
\end{figure}

Current techniques of synthesizing 3D scenes include selecting a set of appropriate objects and generating plausible layouts for the objects. We do ``layout" while techniques for selecting objects are easily incorporated such as \cite{he2020style}. Note that in this paper, we prefer instance-based priors to category-based priors, e.g., we consider a spatial relation between a specific coffee table to a specific chair, where both of them have unique textures and geometries. If categories are being based, distinct features of objects are lost. As shown in Fig. \ref{fig:whyInstance}, different shapes of several armchairs intuitively have their own priors to the same coffee table. 

In this paper, we make the following contributions: 
\begin{enumerate}
    \item We first introduce and learn hyper-relations among three or more objects, which increases layouts of objects of same coherent groups and requires $O(1)$ time to sample layouts of each group, e.g., a coffee table surrounded with several distinct sofas and a TV stand, thus increasing the overall performance. 
    \item We propose a new scaleble geometry-based framework for layout generation, which considers detailed aspects of room layout, e.g., doors, windows, wall decorations, small objects, etc. In coordination with hyper-relations, more plausible and robust layouts are generated. 
    \item We develop an open-source platform for manipulating 3D scenes, where requirements such as rendering, exploring, modifying, etc, are supported, thus allowing researches focus thoroughly on algorithms and techniques. 
\end{enumerate}

\section{Related Works}
3D scene synthesis is to select a set of appropriate objects and transform them plausibly \cite{asurveyof3didss}. Earlier works of synthesizing 3D scenes are mainly based on designing rules, e.g., \cite{germer2009procedural,lyons2008ten} (not data-driven) or data-driven priors. For the former, designing rules are mathematically formulated as a set of constraints followed by optimizations \cite{merrell2011interactive,weiss2018fast}. For the later, since learnt distributions are too complicated to be differentiated, MCMC is assembled for attempting proposals of solving such situation \cite{yu2011make,liang2017automatic,liang2018knowledge,qi2018human,yeh2012synthesizing}. 

Some of them present a framework including both object selection and layout generation, while the rest focus on layouts, though it may also focus on selecting objects \cite{he2020style}. Our method focuses on generations of layouts, i.e., we contribute mainly on how to make layouts more plausible and robust. 

With the continuous study of neural network, several works based on convolutional or graph neural network are proposed \cite{wang2018deep,wang2019planit,ritchie2019fast}, including the current state-of-art work PlanIT \cite{wang2019planit} which is the baseline of our framework. \sk{One} feature of network-based works is \sk{that} they couple selections and layouts, i.e., selecting an object depends on pending layouts, vice versa. In contrast earlier works aforementioned seperate two stages. \sk{Literature based on other techniques does exists, e.g., human-centric assessments \cite{fu2020human}}. Please refer to a more insightful survey on 3D indoor scene synthesis \cite{asurveyof3didss}. 

Several works also synthesize 3D scenes with input other than 3D scenes. Xu et al. recovery 3D scenes from hand sketch \cite{xu2013sketch2scene}. Luo et al \cite{luo2020end} generate 3D scenes from scene graphs. \cite{avetisyan2018scan2cad,chen2014automatic,shao2012interactive} generate room layouts based on RGB-D images or 3D scans. \cite{xiong2020motion,fisher2012example} generate scenes based on input examples. \cite{chang2015text,ma2018language} translate human language to 3D scene configurations. However, different input results in different constraints, frameworks and even applications, these works are beyond the scope of this paper. 

\section{Definitions} \label{sec:def}
Given a list of objects with doors, windows and a room shape, we formulate its corresponding graph $G=<V,E>$ where each object $o \in V$. $E$ is the set of edges which are also simply relations between/among objects. Note that in this paper, we assume a relation may involve more than two instances i.e., a hyper-relation among objects (section \ref{sec:priors}). 

\begin{figure}
    \centering
    \begin{subfigure}[b]{0.495\linewidth}
		\includegraphics[width=\linewidth, height=2.5cm]{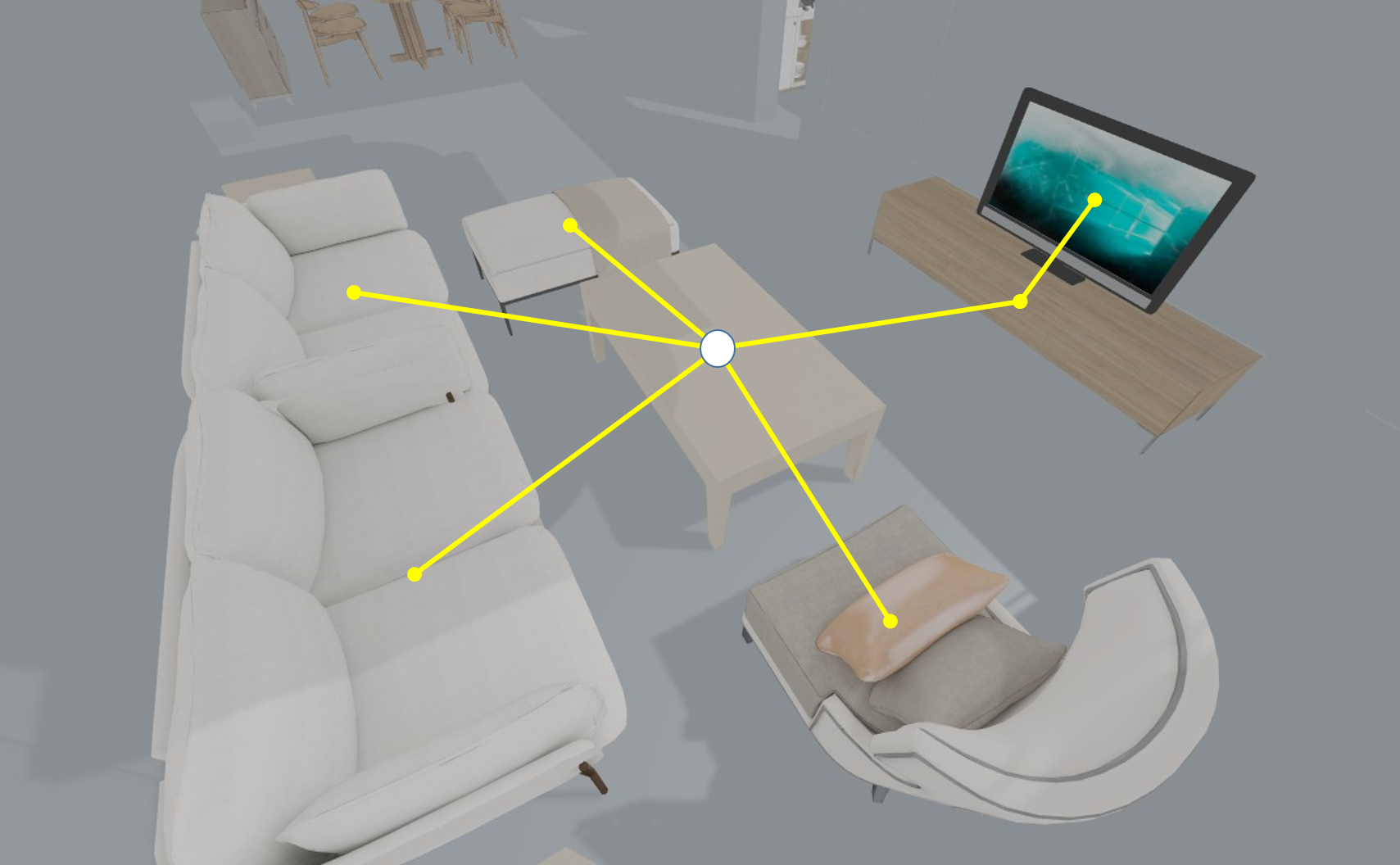}
	\end{subfigure}
	\begin{subfigure}[b]{0.495\linewidth}
		\includegraphics[width=\linewidth, height=2.5cm]{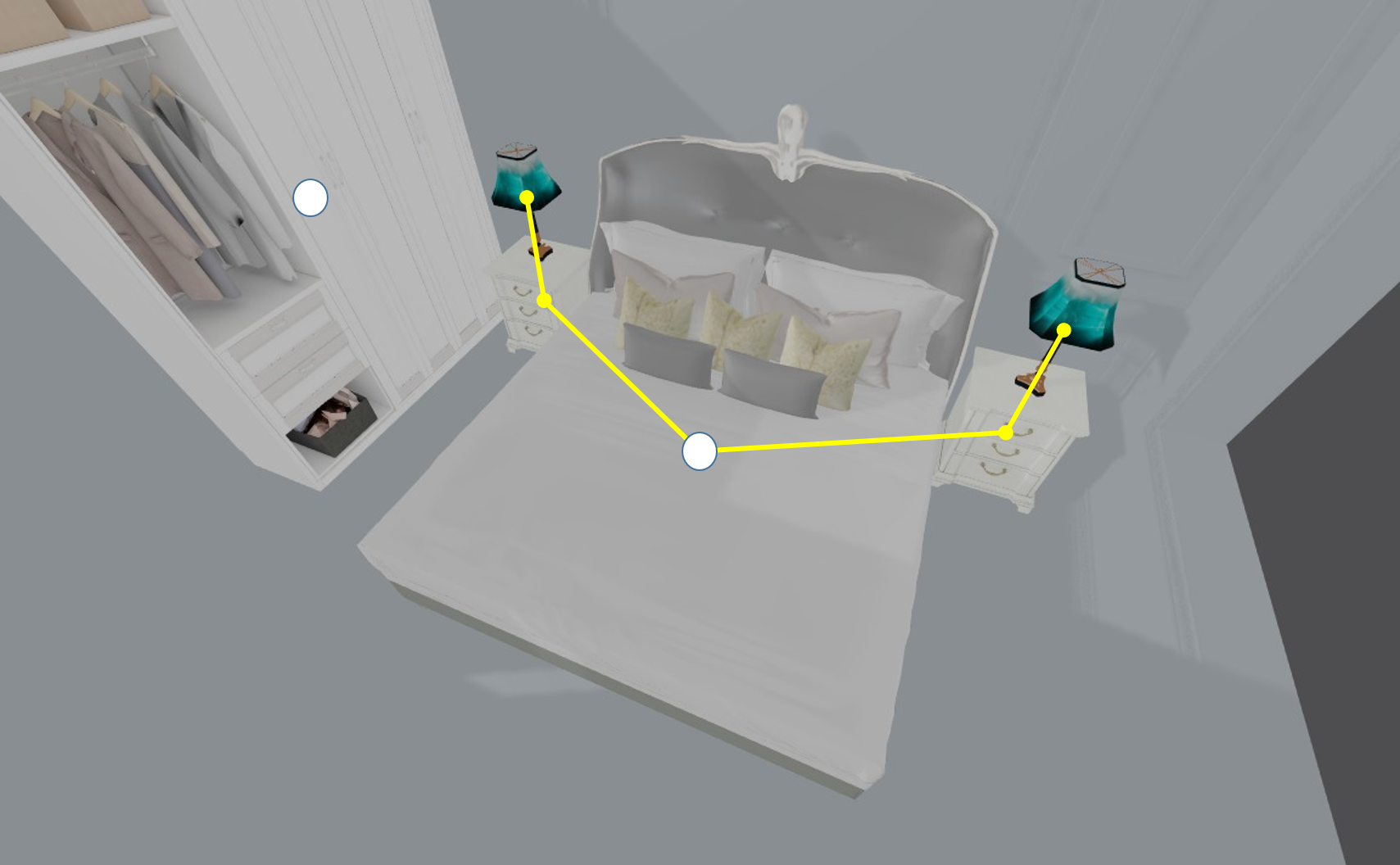}
	\end{subfigure}
    \caption{Three coherent groups where white dots denote respective dominant objects. }
    \label{fig:coherentgroups}
\end{figure}

A coherent group $g$ is a list containing objects where one object connects at least to one another object in the same group. In other words, two coherent groups never have an edge between their instances respectively. Conceptually, generating coherent groups $g \in G$ is equivalent to formulating maximal connected subgraph of $G$, given priors as connections. When generating layout given input, we always initially group objects into several $g_i \subset V$ even though a group may contain one object, such as a wardrobe, a picture frame, a kitchen cabinet, etc. Coherent groups are hierarchical as shown in Fig. \ref{fig:coherentgroups}, where visual edges are pairwise between parents and children.


A transformation of an object includes its translation $(x, y, z)$ and Y-axis rotation $\theta$ where floors align with XoZ plain. In this paper, we do not re-scale objects. The same is true of coherent groups. Additionally, transformations of coherent groups are propagated to their subordinate objects. 

Priors are used to group objects into coherent groups and suggest layouts within each coherent group. A prior set $P_{o_1, o_2, o_3, ..., o_n}$, abbreviated as $P_O$, involves two or more objects. A single prior $p_{O}^{k} \in P_{O}$ suggests a set of plausible transformations for all objects involved. Each prior set contains a dominant object such as $o_1$, and other secondary objects. For example, if a dinning table is surrounded with several chairs and supports a plant, ``dinning table" is the dominant object in this scenario and remaining objects are secondary objects. If only two objects are involved in $P_O$, $P_O$ is a ``pairwise relation" between the two objects (section \ref{sec:pairwise}). If more than two objects are involved and all secondary objects derives from the same instance, $P_O$ is a ``pattern chain set". Otherwise, $P_O$ is a ``hyper-relation" \ref{sec:hyper}. 



\section{Priors} \label{sec:priors}
\begin{figure}
    \centering
	\begin{subfigure}[b]{0.325\linewidth}
		\includegraphics[width=\linewidth]{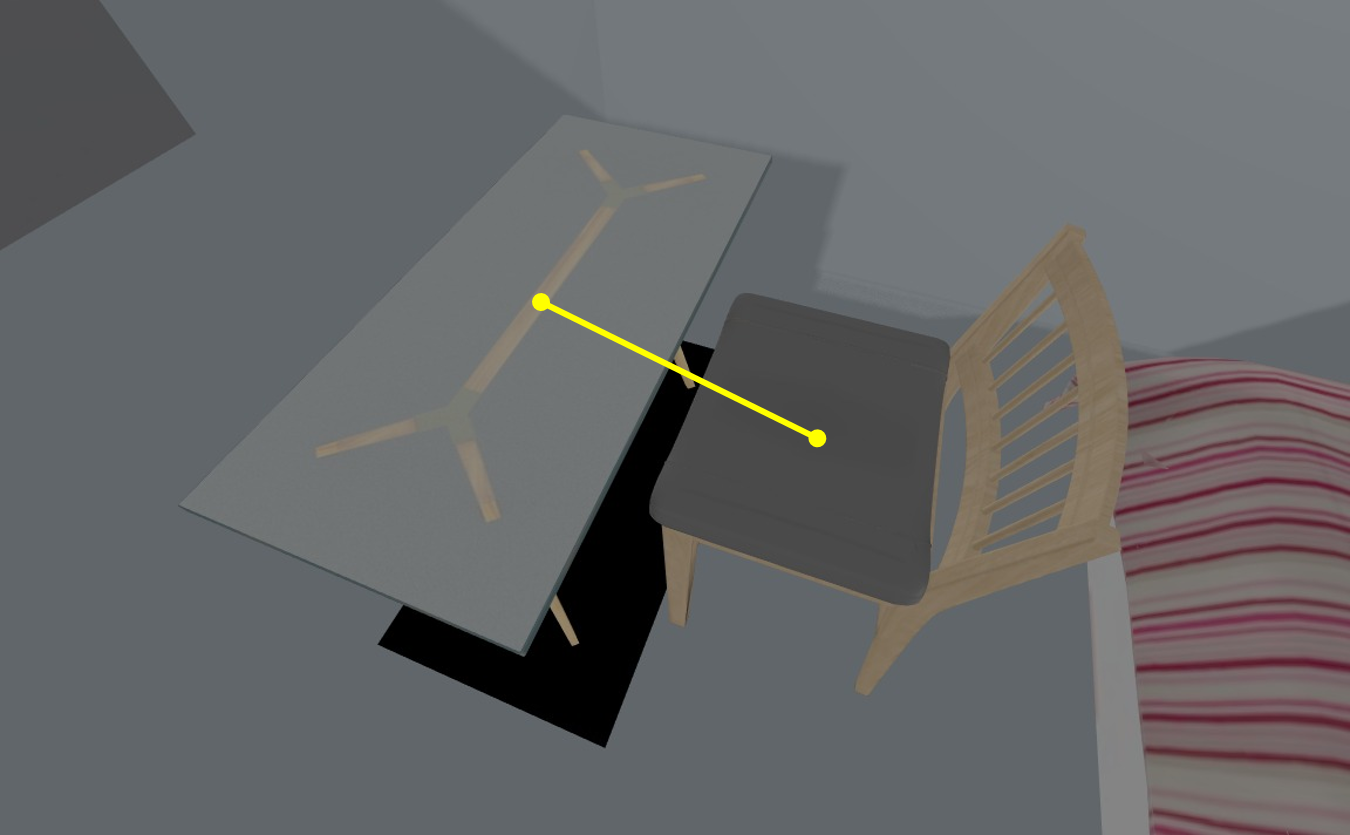}
		\caption{\scriptsize Pairwise Relation. }
		\label{fig:pairwisedemo}
	\end{subfigure}
	\hfill
	\begin{subfigure}[b]{0.325\linewidth}
		\includegraphics[width=\linewidth]{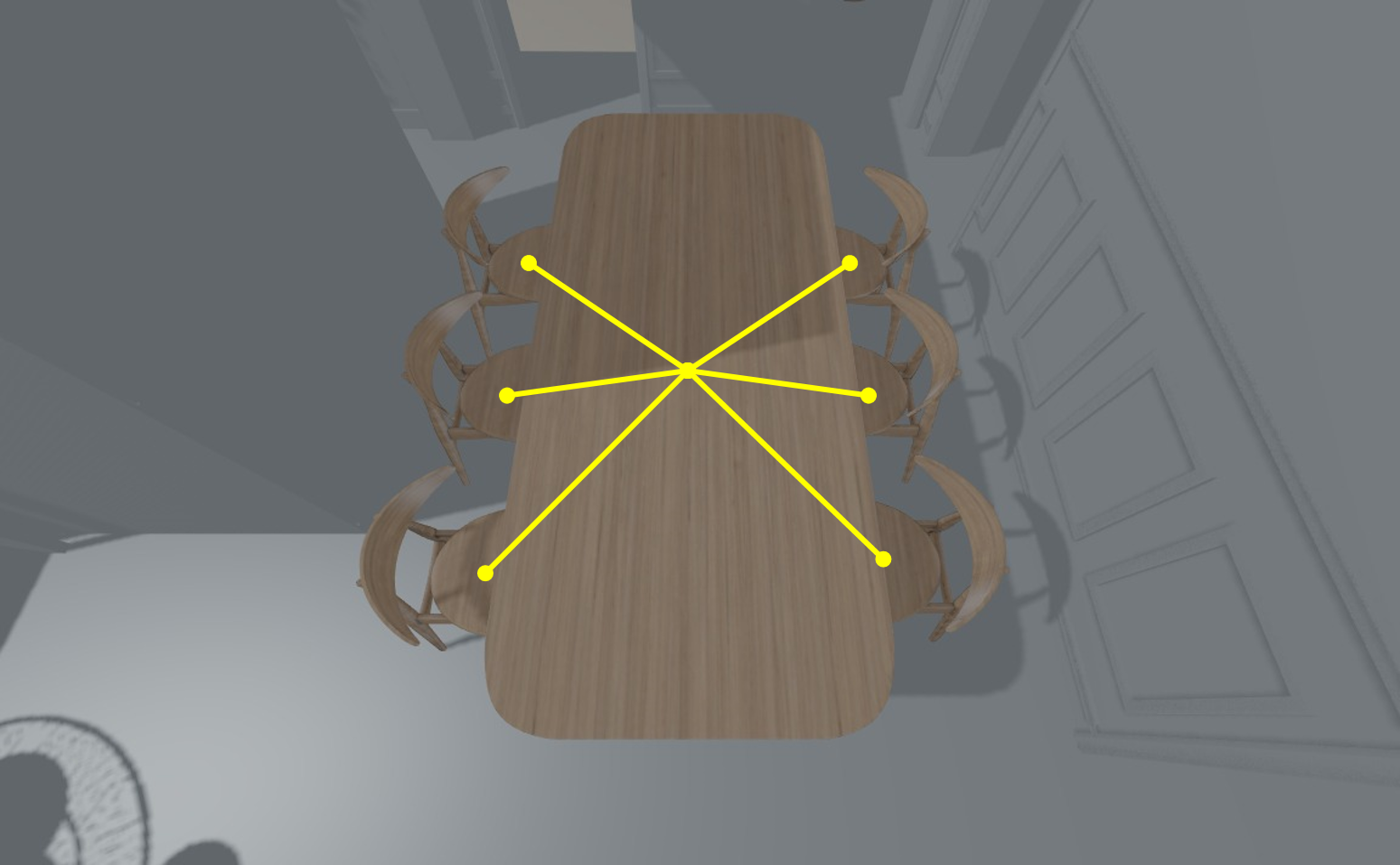}
		\caption{\scriptsize Pattern Chain. }
		\label{fig:chaindemo}
	\end{subfigure}
	\hfill
	\begin{subfigure}[b]{0.325\linewidth}
		\includegraphics[width=\linewidth]{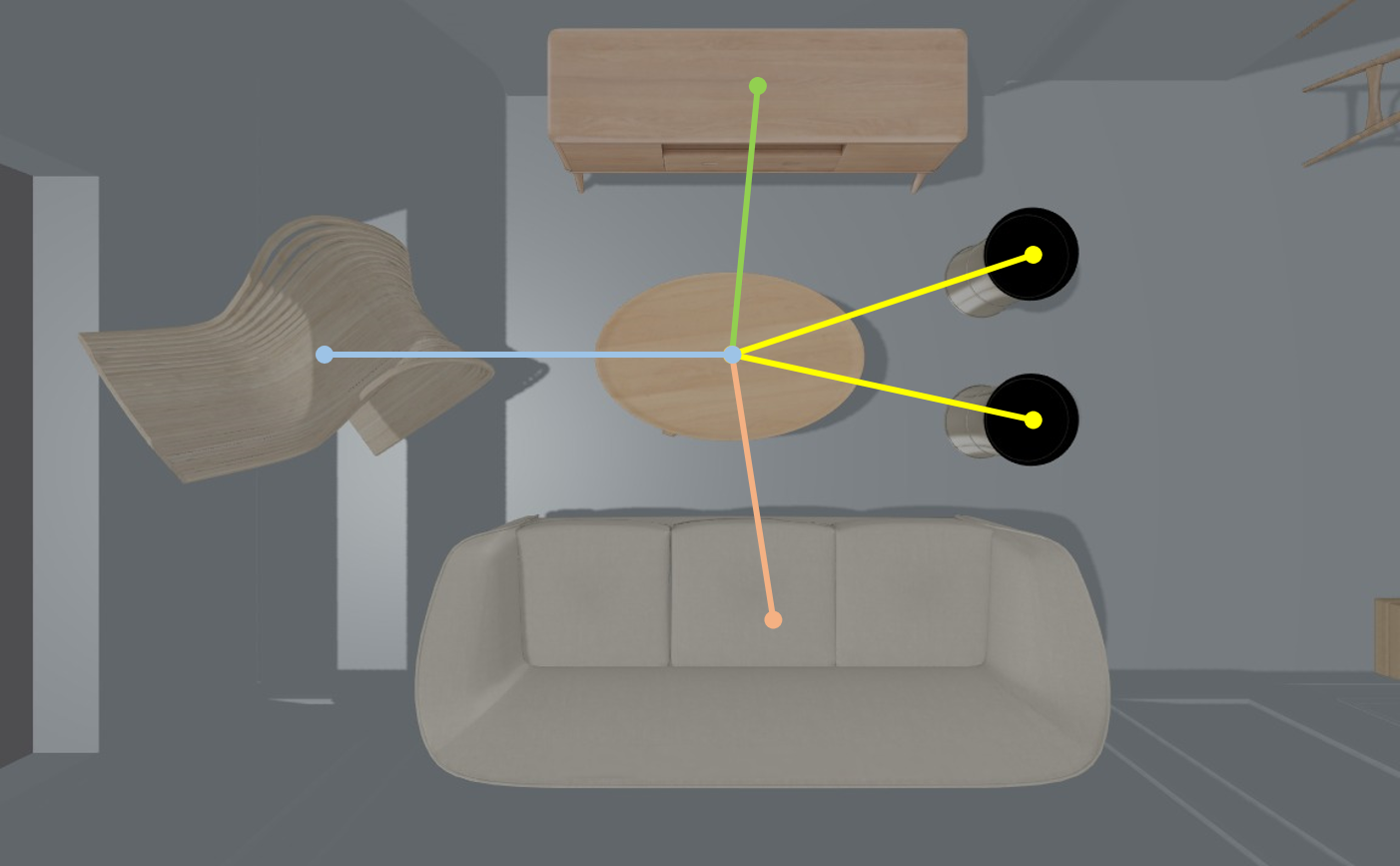}
		\caption{\scriptsize Hyper-Relation. }
		\label{fig:hyperdemo}
	\end{subfigure}
    \caption{Three types of priors in this paper. Links with the same color suggest same secondary objects. \ref{fig:pairwisedemo}: a pairwise ``one-to-one" relation between a desk and a chair; \ref{fig:chaindemo}: a pairwise ``one-to-many" relation between a table and several identical chairs; \ref{fig:hyperdemo}: a hyper-relation among several different objects dominated by a coffee table. }
    \label{fig:threepriors}
\end{figure}

In this section we show how we extract relations among objects. We start by extracting traditional pairwise relations, e.g,. a desk with respect to a chair. Then, we present pre-computed pattern chains which generalize one-to-one relations to one-to-many relations, e.g., a dining table surrounded by several identical chairs. Finally, based on pairwise relations and pattern chains, we further generalize and formulate hyper-relations among objects, i.e, a relations ``among" more than two instances. Fig. \ref{fig:threepriors} suggests the differences of them. In this section, we show how priors are represented and generated, and the usage is shown in section \ref{sec:layoutGen}. 

Theoretically, pairwise relations and pattern chains are both special forms of hyper-relations. The reason of discussing them separately is because directly learning hyper-relations is difficult. As a result, we first introduce pairwise relations which derive more general pattern chains thus enabling forming hyper-relations. 

\subsection{Pairwise Relation} \label{sec:pairwise}
A Pairwise relation is a set of priors $P_{ab}$ from a dominant object $a$ to a secondary object $b$. Given a pairwise relation $P_{ab}$, we can sample a prior $p_{ab,k} \in P_{ab}$ that is directly a transformation of $b$ with respect to $a$. Note that pairwise relations are directional, and sampled transformations are only relative between two objects involved i.e., global transformations are still required (section \ref{sec:layoutGen}). 

We extract discrete pairwise priors by utilizing density peak clustering (DPC) \cite{rodriguez2014clustering}, which firstly calculate $\rho_{k} = \sum_{k'} I_{\{d \leq d_{c}\}}(d_{k,k'}), d_{c} = d_{(0.015 K^2)}$ and $\delta_{k} = \min_{k':\rho_{k} < \rho_{k'}}(d_{k,k'})$ for all points. \sk{In our situation, $d_{k,k'}$ denoting the Euclidean distance from the transformation of dominant object $k$ to the transformation of secondary object $k'$. A transformation includes transitions and rotations. $d_c$ is a hyper-parameter and $rho_{k}$ is counting the number of $d_{k,k'}$ that is lower than $d_c$. The selection of $d_c$ follows \cite{rodriguez2014clustering}, i.e., the $0.015 K^2$th greatest $d_{k,k'}$ among all $k^2$ relative distances. $\delta_{k}$ seeks a minimal $d_{k,k'}$ among all $d_{k,k'}$ with higher $rho_{k'}$ than $rho_{k}$.} Please refer to \cite{rodriguez2014clustering} for more details about this algorithm. Although DPC is typically used for clustering, it does anomaly detection for eliminating noises, i.e., removing points with low values of $\rho$ and high values of $\delta$. Cluster centers and ordinary points are treated equally since they are already reasonable transformations in this paper.

After elimination, remaining ``points" are plausible relations directly from datasets (human designers) where each ``point" become a single pairwise prior $p_{ab,k} \in P_{ab}$ for locally arranging a dominant object and its secondary object. Typical dominant objects include desk, dinning table, coffee table, bed, etc. We manually label a set of instances that are capable of being dominant objects.

\subsection{Pre-Computed Pattern Chain}

\begin{figure}
    \centering
	\begin{subfigure}[b]{0.32\linewidth}
		\includegraphics[width=\linewidth, height=2.5cm]{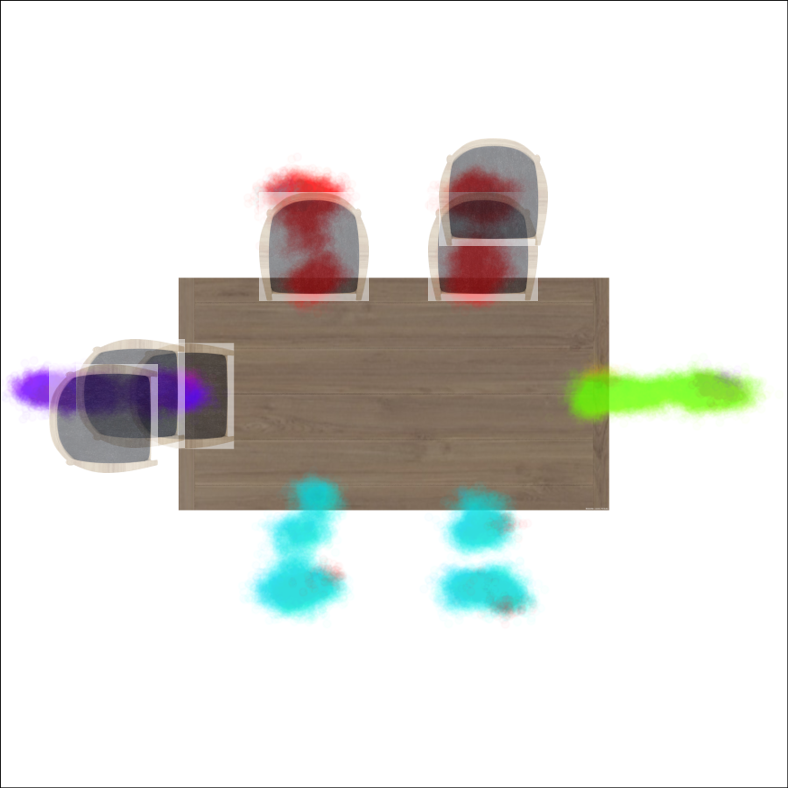}
		\caption{}
		\label{fig:patternChain1}
	\end{subfigure}
	\hfill
	\begin{subfigure}[b]{0.32\linewidth}
		\includegraphics[width=\linewidth, height=2.5cm]{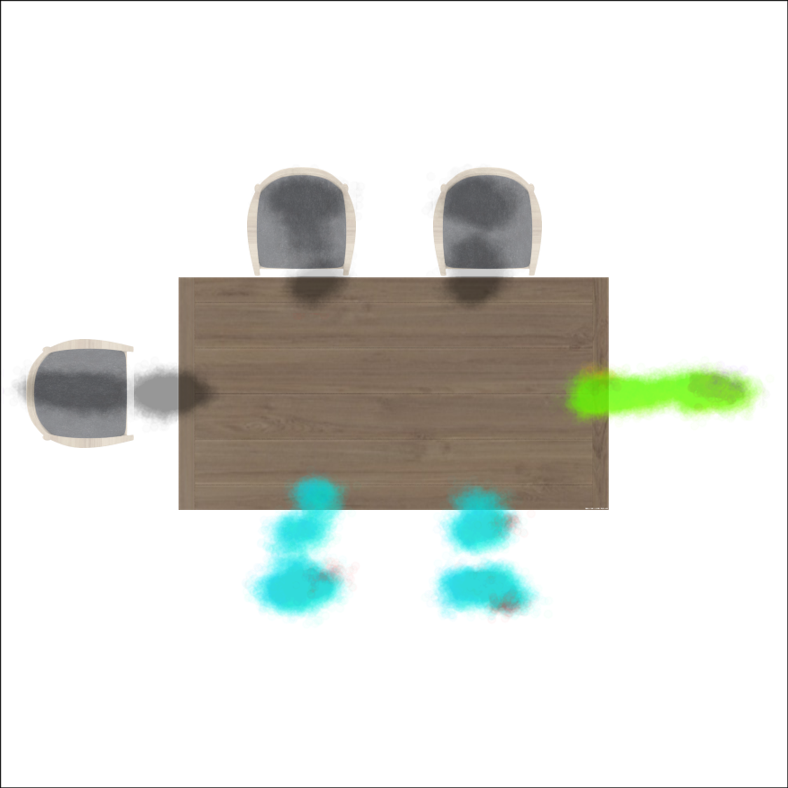}
		\caption{}
		\label{fig:patternChain2}
	\end{subfigure}
	\hfill
	\begin{subfigure}[b]{0.32\linewidth}
		\includegraphics[width=\linewidth, height=2.5cm]{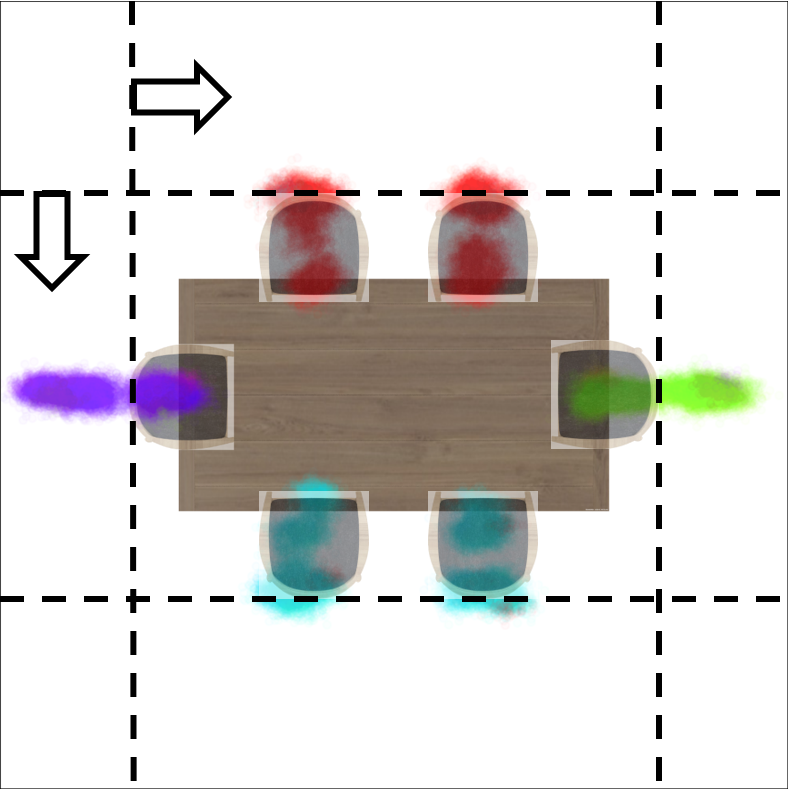}
		\caption{}
		\label{fig:patternChain3}
	\end{subfigure}
    \caption{\ref{fig:patternChain1}: Directly sampling a pairwise relation without pre-computed pattern results in obvious implausibility. \ref{fig:patternChain2}: Recursively formulating a pattern chain. \ref{fig:patternChain3}: Additional constraints are optional if e.g., well-aligned layouts are required. }
    \label{fig:patternchain}
\end{figure}

Commonly, a dominant object has several secondary copies of the same instance, e.g., a dinning table with several identical chairs. If we sample them twice or more as shown in figure \ref{fig:patternChain1}, aforementioned pairwise relations do not guarantee plausibility of ``one-to-many" relations. Thus, we solve it by presenting pattern chains. 

A pattern chain set $C_{ab}$ is a prior set between object $a$ and $b$. \sk{Each $c_{ab}^{j} = \{ j_1, j_2, ..., j_n \}, c_{ab}^{j} \subset \mathbb{N}$ is a list of indices to its pairwise relation $P_{ab}$, e.g., $j_x$ indexes to the $x$-th pairwise relation $p_{ab,j_x}$ in $P_{ab}$}. Generating one pattern chain $c_{ab}^{j}$ is a recursive process. First, a $p_{ab,j_1} \in P_{ab}$ is randomly selected from $P_{ab}$. As discussed, $p_{ab,j_1}$ gives a plausible transformation between $a$ and $b$. Second, we traverse all $p_{ab,i} \in P_{ab}$. If a copy of object $b$ with the transformation of $p_{ab,i}$ do not collide with another copy with the transformation of $p_{ab,j_1}$, $p_{ab,i}$ is included in a new subset $P'_{ab} \subset P_{ab}$. Third, we would like to place another copy of $b$, so $p_{ab,j_2}$ is randomly selected from $P'_{ab}$ and the above procedure is executed recursively until $P'_{ab}$ is empty. As shown figure \ref{fig:patternChain2}, after three iterations, placing three chairs around a table filters out a subset of their pairwise priors (gray). Therefore, a fourth chair can be only placed in the remaining pigmented areas. When a chain is generated, we can optionally adjust it, e.g., figure \ref{fig:patternChain3} suggests ``horizontals and verticals" to make the chain well-aligned. 

Note the above generates one pattern chain $c_{ab}^{j} = \{ j_1, j_2,... \}$. In theory, a $P_{ab}$ of $O(n)$ size has $O(n!)$ undetermined pattern chains. In practice, we only generate one pattern chain for each $p_{ab,k} \in P_{ab}$, to make sure each pairwise relation is used at least once, instead of figuring out the entire pattern chain set. Otherwise, it requires $O(n!)$ time and space to compute only a single set, which also slow down online arrangement by restricting prior loading. 

\subsection{Hyper-Relation} \label{sec:hyper}
\begin{figure}
    \centering
	\begin{subfigure}[b]{0.325\linewidth}
		\includegraphics[width=\linewidth, height=2cm]{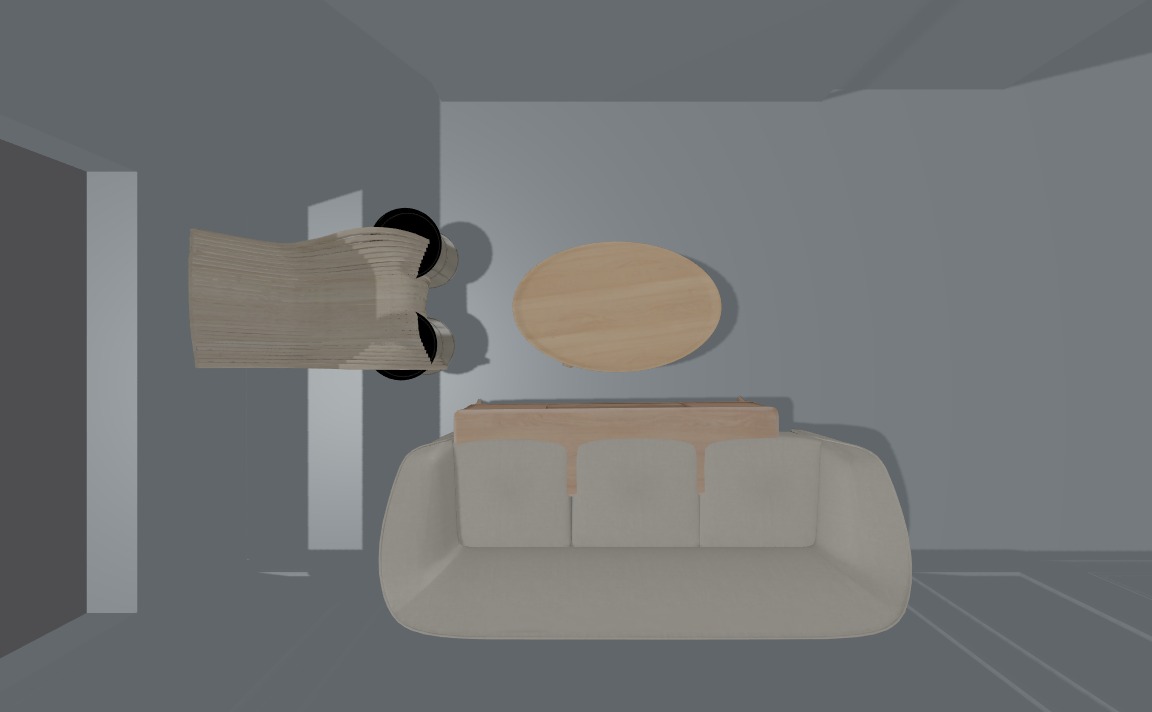}
		\caption{}
		\label{fig:hyper1}
	\end{subfigure}
	\hfill
	\begin{subfigure}[b]{0.325\linewidth}
		\includegraphics[width=\linewidth, height=2cm]{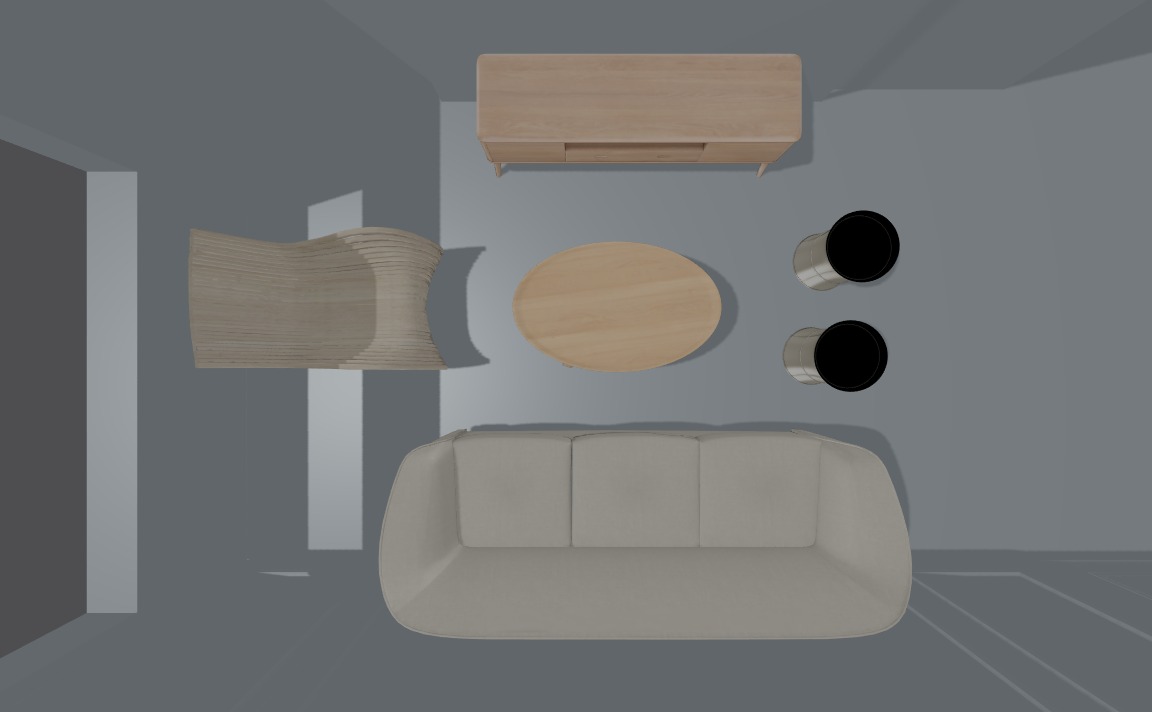}
		\caption{}
		\label{fig:hyper2}
	\end{subfigure}
	\hfill
	\begin{subfigure}[b]{0.325\linewidth}
		\includegraphics[width=\linewidth, height=2cm]{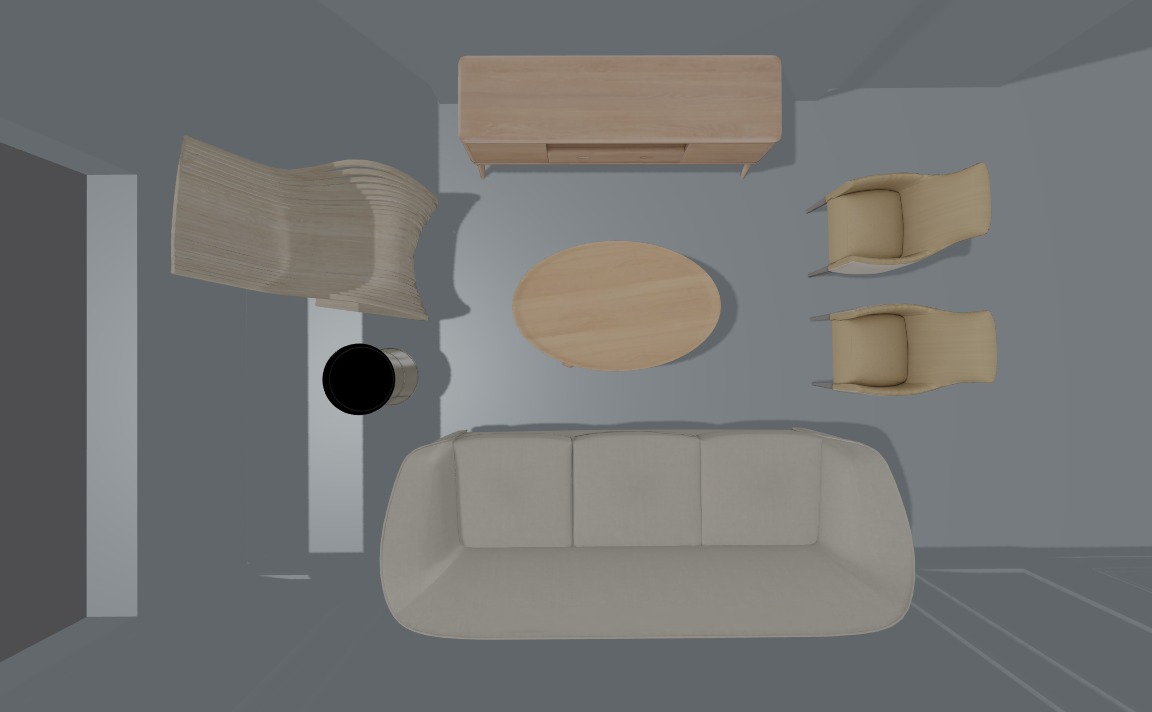}
		\caption{}
		\label{fig:hyper3}
	\end{subfigure}
    \caption{\ref{fig:hyper1}: Using only pairwise relations results implausibilities among secondary objects in a coherent group. \ref{fig:hyper2}: Using hyper-relations results possibilities among all objects involved. \ref{fig:hyper3}: A different object set requires another hyper-relation, since we can not assume ``as many objects as possible". }
    \label{fig:hypermain}
\end{figure}

A hyper-relation $H_O$ is a prior set among several objects $O=\{o_{dom}, o_{sec1}, o_{sec2}, ... \}$. A dominant object $o_{dom}$ exists in $H_O$ such as a coffee table and secondary objects relate to each other, e.g., chairs on a rug, armchairs beside a long sofa. Purely sampling pairwise prior sets results in scenarios such as figure \ref{fig:hyper1}, where secondary objects are only plausible with respect to their dominant object. Hyper-relation is essentially different from pattern chains. Pattern chain sets are still one-to-one relations and a pattern chain assumes incorporating as many secondary objects as possible. In contrast, a hyper-relation has a definite list of objects, i.e., we can not assume what instances are included and how many copies each instance has in a specific hyper-relation, because areas are limited. As shown in figure \ref{fig:hyper2} and \ref{fig:hyper3}, different numbers and instances of seats derives two distinct hyper-relations. 


To generate hyper-relations, we do not hypothesize and learn concrete distributions because real distributions are too complicated to be expressed, solved and sampled \cite{li2018df}. Instead, we try achieving as many exact samples as possible. Given a set of objects $O$ and its dominant object $o_{dom} \in O$, we randomly select a secondary object $o_{sec} \in O$ and randomly sample a prior from the pairwise relation between $o_{dom}$ and $o_{sec}$. Thus, $o_{sec}$ is transformed with respect to $o_{dom}$. Next, similar to generating pattern chains, we filter the remaining pairwise relations between $o_{dom}$ and other secondary objects $o_{sec1}, o_{sec2}, ... \in O$, to ensure ``collision free". With multiple instances, additional rules are required. We use ``tiers", which as far as we studied is firstly terminologized in \cite{yu2011make}, for finer filtering. For example, rugs are placed on the ground where objects such as tables and beds can be put on top of it. Merely detecting collisions would mistakenly filter plausible priors. Not detecting collisions between objects of different tiers alleviates such situations. After filtering the remaining pairwise relations, recursively, we randomly select another secondary object and repeat the above steps until all secondary objects are placed appropriately with no implausibilities. After that, a single hyper-prior is generated with transformations of all secondary objects. We iteratively re-run the entire process to enrich the pending hyper-relation. 


Yet, the above steps still require definite lists of objects. Nevertheless, figuring out all undetermined lists is almost equivalent to exhaustively traverse all combinations of objects. To address this, we systematically optimize extractions. After forming coherent groups (section \ref{sec:layoutGen}), we check their hierarchies. If a parent has two or more children, we try assemble the hyper-relation for them. If the hyper-relation does not exist, a new thread is started to generate it in background. \sk{In other words, we either load existing hyper-relations if they are already generated or establish a thread for generating them when we need them.} Alternatively, users can manually suggest their own lists of objects to generate their hyper-relation. 


\section{Geometry-Based Layout Generation} \label{sec:layoutGen}
\subsection{Coherent Grouping}
\begin{figure}
    \centering
    \includegraphics[width=\linewidth]{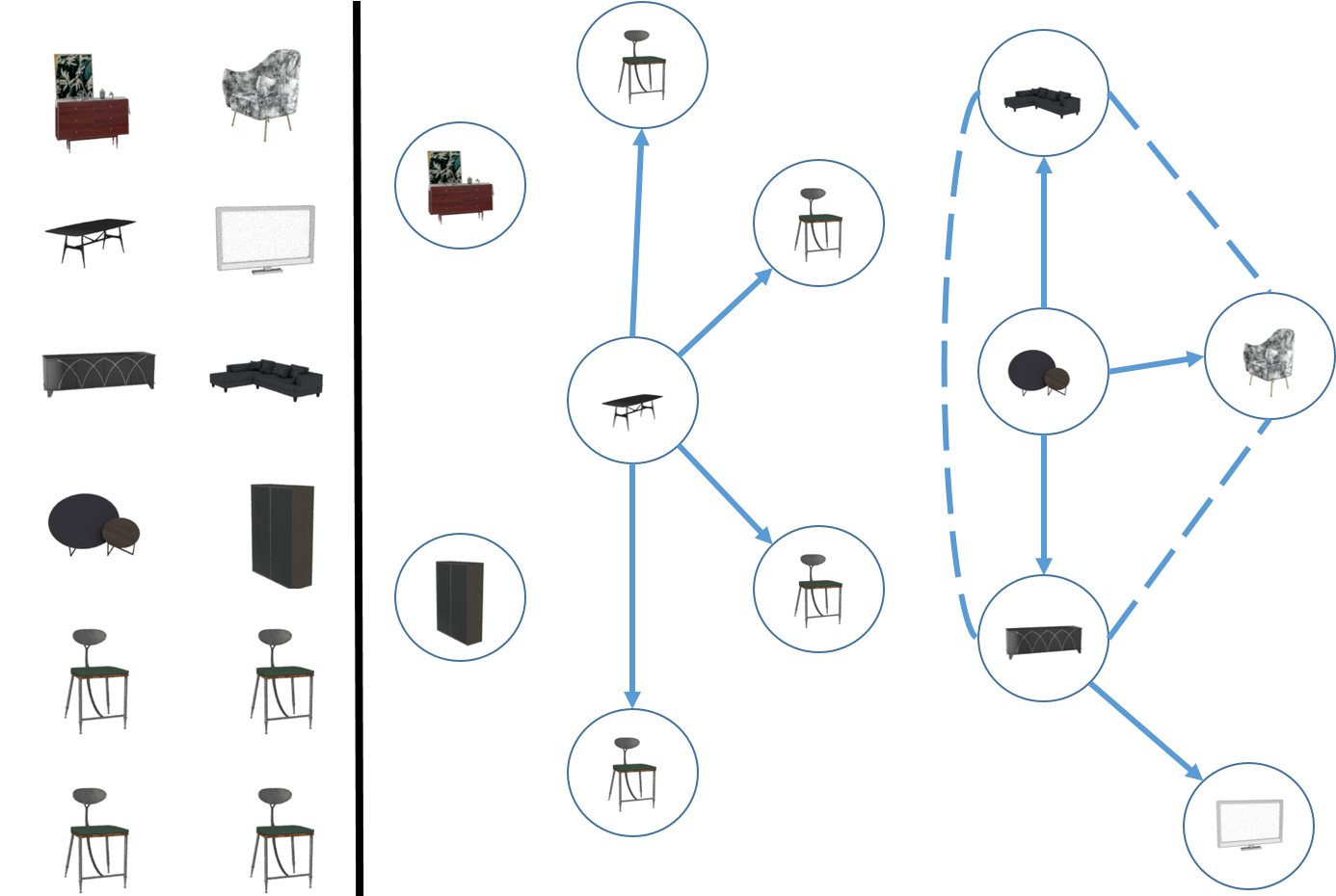}
    \caption{Coherent Grouping. Dotted dashes denote hyper-relations of secondary objects. Given a list of objects to generate their layout, we first group them into several coherent groups. For example, a coffee table relates to two sofas and a TV stand and the TV stand relates to a TV, so they form one coherent group. Two cabinets have no relation to others, so each of them form their own groups. }
    \label{fig:coherentgrouping}
\end{figure}

We show how we finally arrange objects in this section. First, objects are decomposed into several coherent groups $g_i \subset G$ based on finding maximal connected subgraphs using pairwise relations between objects as shown in figure \ref{fig:coherentgrouping}, where whether or not two objects are connected depends on existence of pairwise relations between objects. 

One secondary object can have at most one dominant object. If multiple available dominant objects exist with respect to a secondary object $o_{sec}$, we randomly select a dominant object and discard relations between $o_sec$ and other dominant objects. Each dominant object also has finite lengths of copies of secondary instances guided by lengths of respective pattern chains. This makes our framework more flexible, e.g., given only one chair but a dressing table and a desk in a bedroom, we randomly assign the chair to either the dressing table or the desk, which gives more variance to generated results. 

After that, input objects are distributed in coherent groups. As discussed in section \ref{sec:priors}, within a specific coherent group, we can directly sample a set of transformations for all objects locally within the group. As shown in figure \ref{fig:coherentgrouping}, if a parent has two or more descendants and each descendants are different, the hyper-relation is assembled or started to be generated in background, e.g., coffee table with respect to two sofas and a TV stand. If the descendants are identical, the pattern-chain set is sampled, e.g., dining table and four chairs. Otherwise, we use pairwise priors, e.g., TV stand and TV. Therefore, the final process is to transform several coherent groups properly in the room. 

\subsection{Geometric Arranging} \label{sec:wallchasing}

Eventually, we assign transformations to each coherent group and propagate transformations to objects. Since priors already layout objects sufficiently within groups, three more constraints are required to make layouts physically plausible among groups: 1, all groups should be inside a room; 2, all groups should not overlap each other; 3, clear paths should exist for windows and doors. 

Placing a set of shapes (coherent groups) in another huger polygon (room) is an np-hard problem \cite{de1997computational} in computational geometry. Thus, we geometrically simplify coherent groups as cuboids, consider doors and windows as fixed (pre-arranged) blocks, and do heuristic attempts as shown in algorithm \ref{alg:ra}. 

\sk{We first sort coherent groups according to their area occupied from the largest to the smallest, since bigger groups usually represent more central functionality of rooms, e.g., a bedroom is call a ``bedroom" due to a coherent group dominated by a bed. Then, coherent groups are placed with regard to this order, whereas random positions are assigned to them along the inner side of the targeting room, since the definition of coherent groups indicates the relations among different coherent groups are weak (section \ref{sec:def}). After placing a group, we check potential collisions between this group and other groups or blocks. If collided, we discard the transformation and randomly re-select a new transformation. To enhance the performance, we used exponentially increasing sampling density. If a proper transformation fails at a density of $d$ for the pending coherent group, we increase $d$ to $2d$ to find more possible positions. But if it still collides after several times of increasing density, we discard the group and conduct the next one. To increase the plausibility, we add more heuristic rules: 1, we initially attempt to transform groups at corners of rooms and sides of other existing coherent groups. During collision detection, we take the height into consideration. So it is possible that some furniture with lower height is placed in front of windows. Finally, ``liftings" $L_f$ are assigned to groups. If $L_f = 0$, a groups is placed against walls. If $L_f$ equals to half the length of the room, a group is placed in the middle of the room.}  

\begin{algorithm}
	\caption{Geometric Arranging} 
	\label{alg:ra}
	\begin{algorithmic}[1]
		\Require ~~\\ 
		Polygon of room's inner side $P_{r} $;\\
		List of rectangles of coherent groups with height $A_{rec}$; \\
		List of rectangles of windows and doors; 
		\Ensure Transformations of rectangles $T_{rec}$;
	    
	   \Function{CheckOK}{$A$}
	        \If{$A$ does not overlap with existing groups and blocks}
	        \State return True
	        \Else    
	        \State return False
	        \EndIf
	    \EndFunction
	    
	    \Function{ApplyTransform}{$A$,$t$}
	        \State apply transformation $t$ to $A$
	        \State return $A$
	    \EndFunction
	       
		\Function {InsertRectangle}{$A$}
		\State Let $T$ be array of transformations
		//For heuristic
		\For{$edge \in P_{r}$ and $p \in $existing polygons}
		\State Push heuristic transformation of $edge$ or $p$ to $T$
		\EndFor
		
		\For{$t \in T$}
		\If{CheckOK(ApplyTransform($A$,$t$))}
		    \State return $t$;
		\EndIf
		\EndFor
			
		//For random
		\State Clear $T$
		\For{$n=1 \to max\,sampling\,density$}
        \For{$edge \in P_{r}$}
		\State Push $2^n*len(edge)$ random transformations on $edge$ to $T$ 
		\EndFor
		
		\State Shuffle $T$
		
		\For{$t \in T$}
		\If{CheckOK(ApplyTransform($A$,$t$))}
		    \State return $t$;
		\EndIf
		\EndFor
		\State Clear $T$
        \EndFor
        \State return None;
		
	    \EndFunction
	    \For{$a \in A_{rec}$}
		\State Push InsertRectangle($a$) to $T_{rec}$;
		\EndFor
		
	\end{algorithmic}
\end{algorithm}

\section{Experiments}

\subsection{Setup}
\begin{figure}
    \centering
    \begin{subfigure}[b]{0.49\linewidth}
		\includegraphics[width=\linewidth]{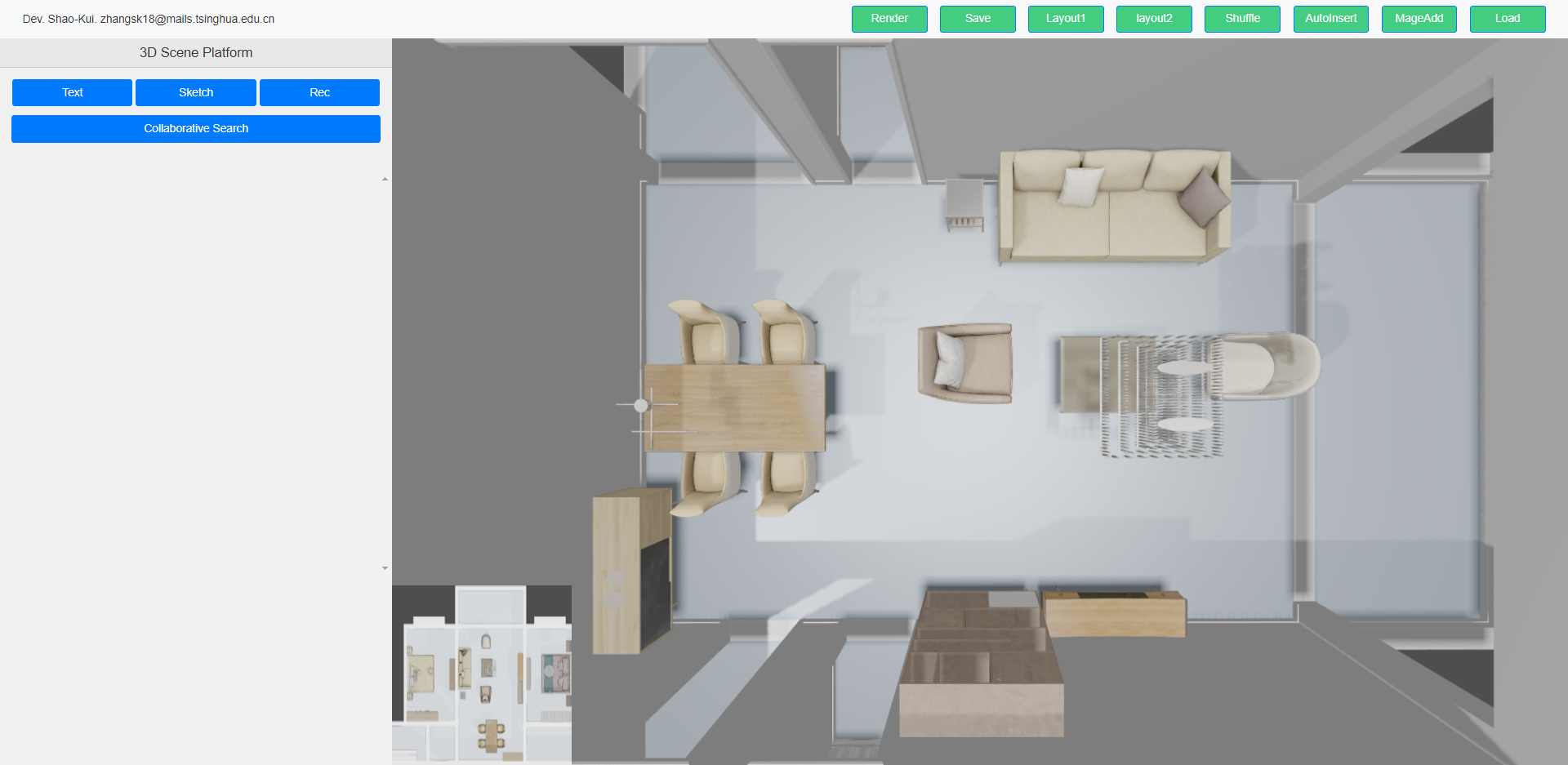}
		\caption{Platform Overview. }
	\end{subfigure}
	\begin{subfigure}[b]{0.49\linewidth}
		\includegraphics[width=\linewidth]{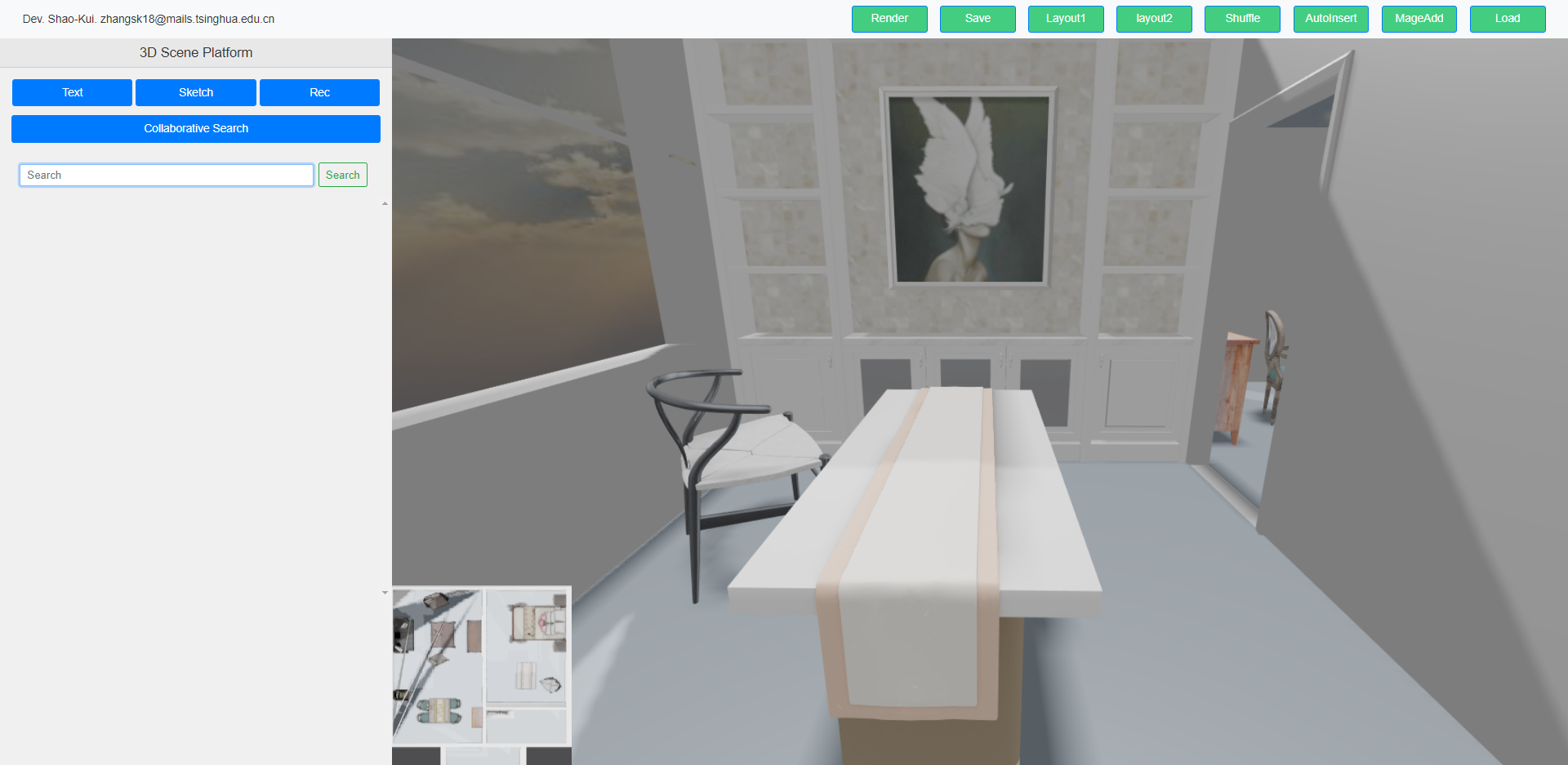}
		\caption{Viewing \& Roaming. }
	\end{subfigure}
	
	\begin{subfigure}[b]{0.49\linewidth}
		\includegraphics[width=\linewidth]{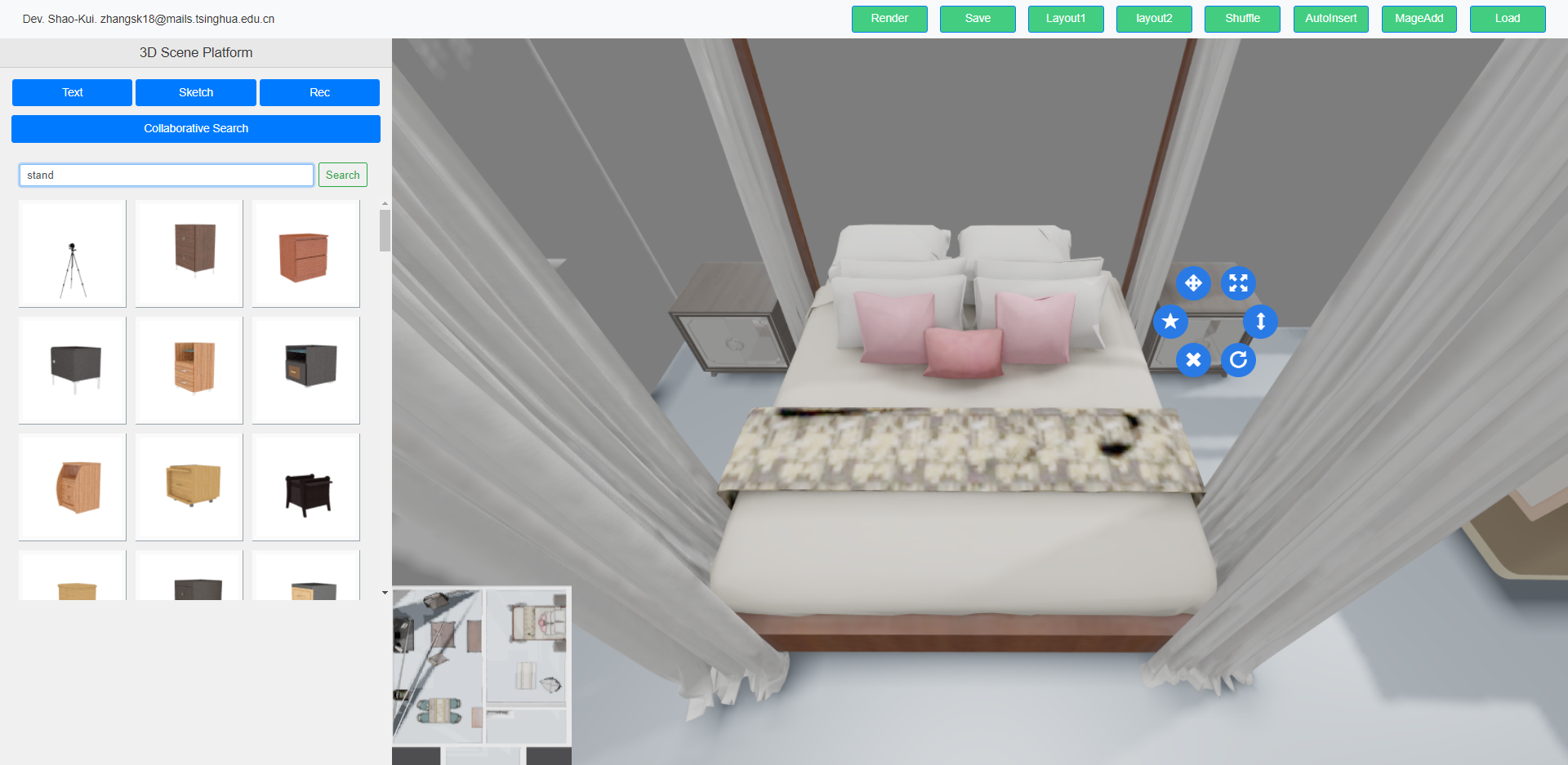}
		\caption{Manipulating \& Searching. }
	\end{subfigure}
	\begin{subfigure}[b]{0.49\linewidth}
		\includegraphics[width=\linewidth]{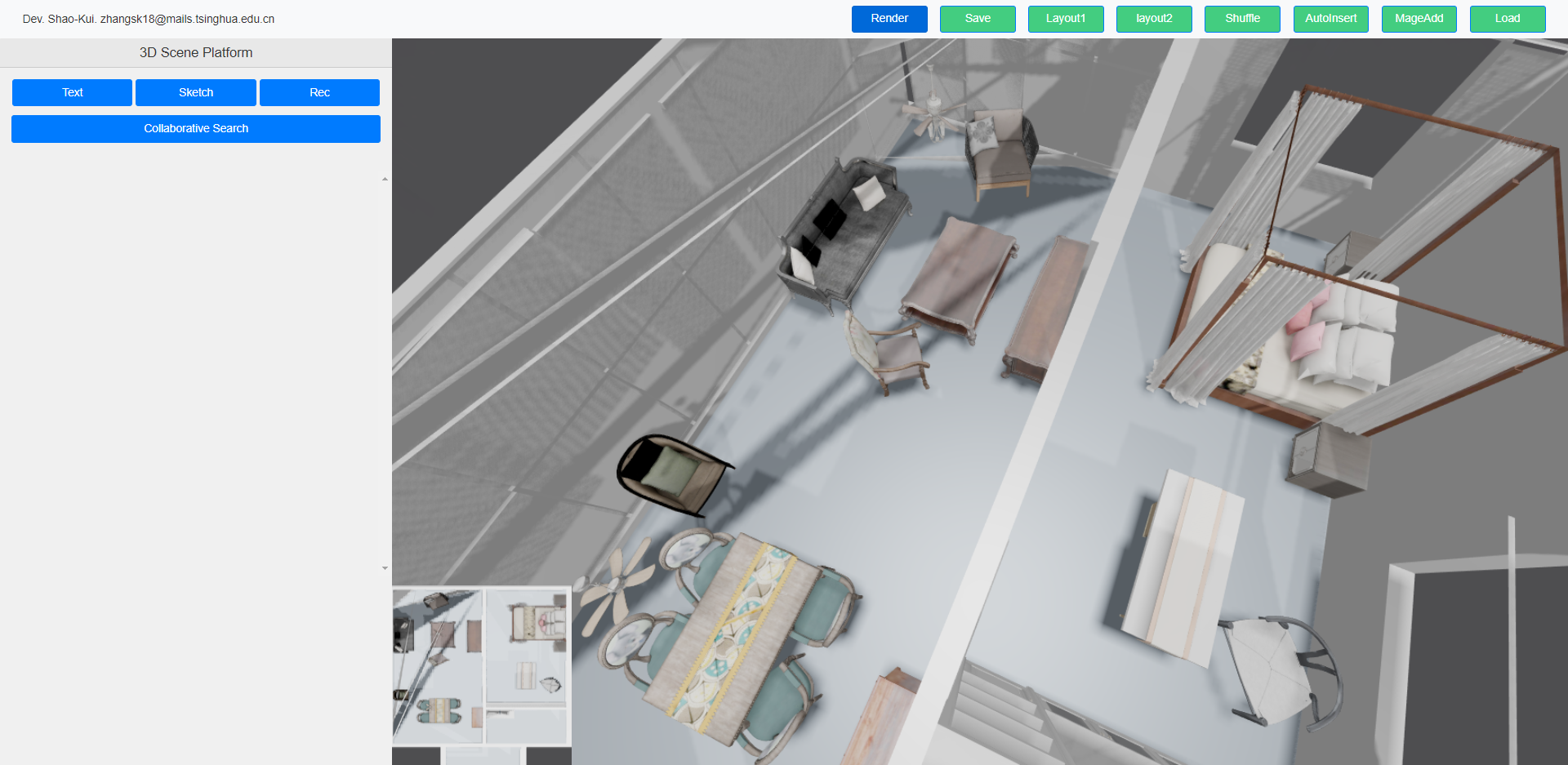}
		\caption{Rendering. }
	\end{subfigure}
    \caption{We develop an open-source 3D scene platform allowing adding, deleting, modifying, searching objects and rendering, saving scenes. Users can explore given 3D scenes by orbital control. Our platform is embedded with the proposed algorithm. }
    \label{fig:platform}
\end{figure}

\begin{figure*}[!h]
	\centering
	
	\begin{subfigure}[b]{0.49\linewidth}
		\includegraphics[width=\linewidth]{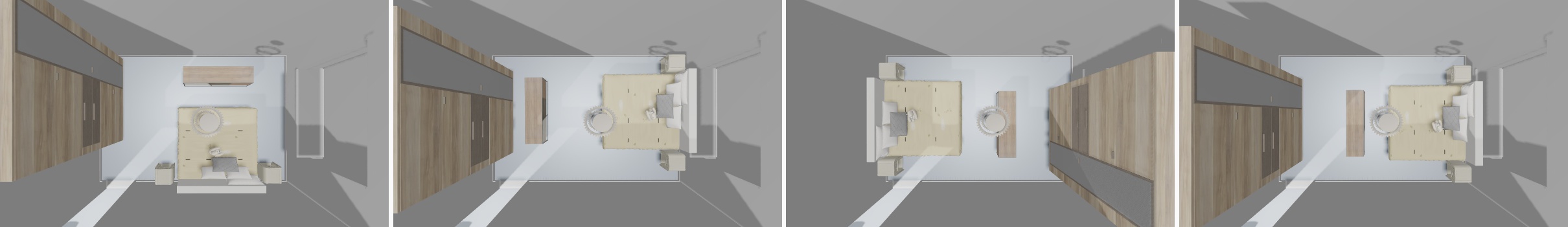}
	\end{subfigure}
	\begin{subfigure}[b]{0.49\linewidth}
		\includegraphics[width=\linewidth]{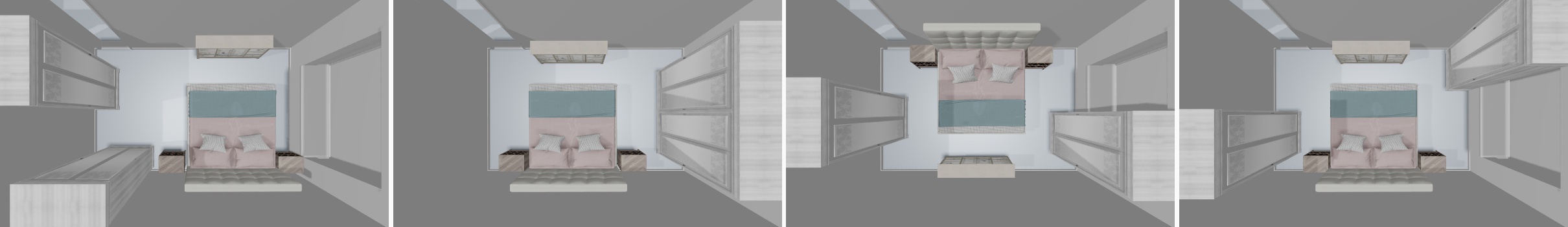}
	\end{subfigure}
	
	\begin{subfigure}[b]{0.49\linewidth}
		\includegraphics[width=\linewidth]{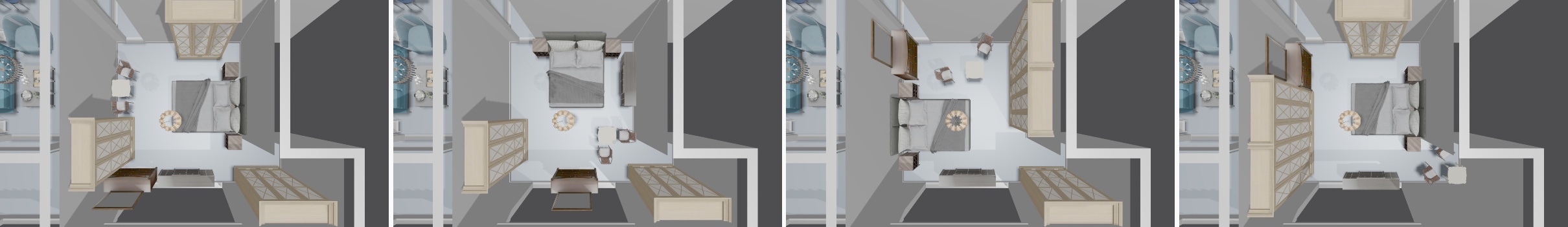}
	\end{subfigure}
	\begin{subfigure}[b]{0.49\linewidth}
		\includegraphics[width=\linewidth]{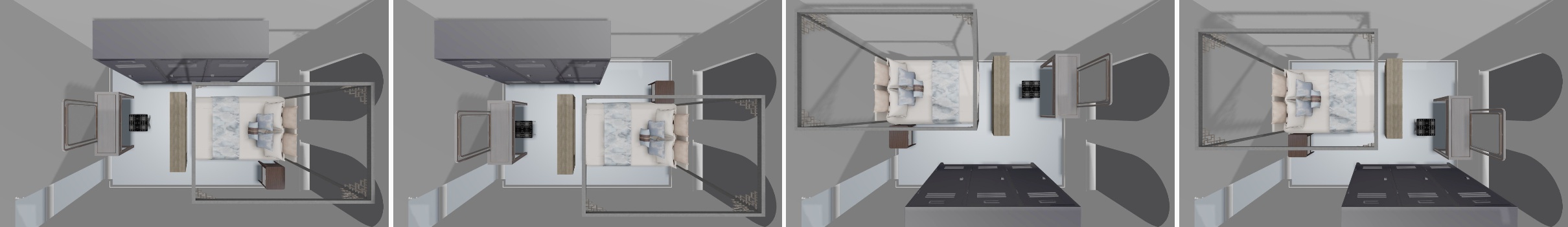}
	\end{subfigure}
	
	\begin{subfigure}[b]{0.49\linewidth}
		\includegraphics[width=\linewidth]{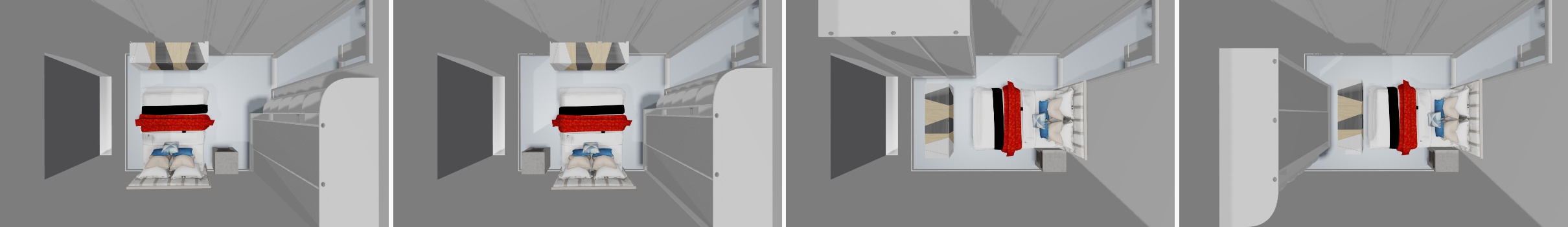}
	\end{subfigure}
	\begin{subfigure}[b]{0.49\linewidth}
		\includegraphics[width=\linewidth]{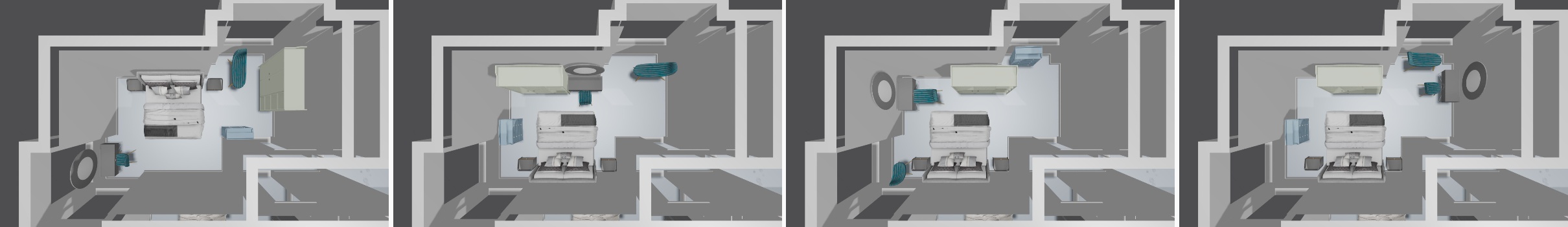}
	\end{subfigure}

	\begin{subfigure}[b]{0.49\linewidth}
		\includegraphics[width=\linewidth]{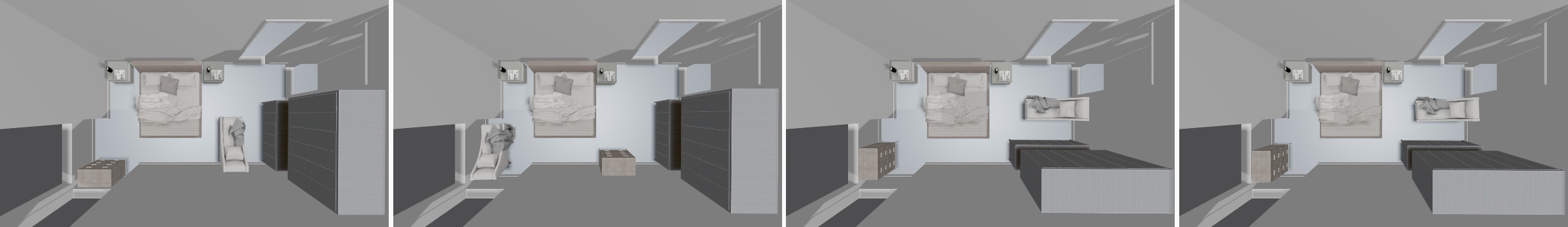}
	\end{subfigure}
	\begin{subfigure}[b]{0.49\linewidth}
		\includegraphics[width=\linewidth]{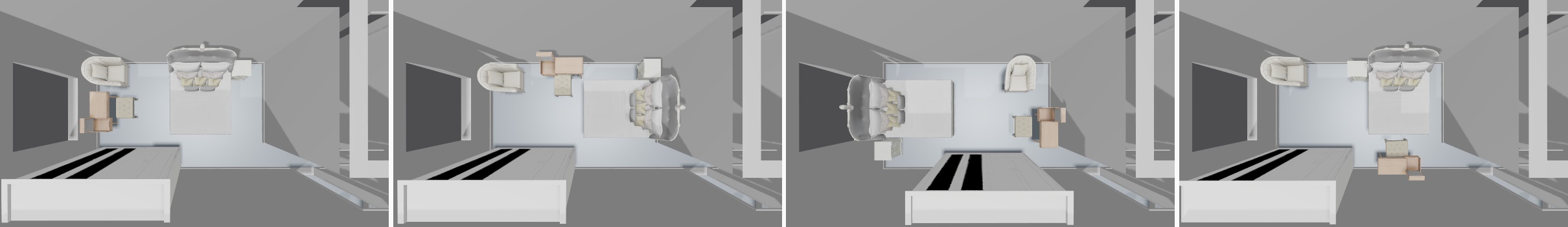}
	\end{subfigure}
	
	\vspace*{-3mm}
	\caption*{Bedroom. }
	\vspace*{2mm}
	
	\begin{subfigure}[b]{0.49\linewidth}
		\includegraphics[width=\linewidth]{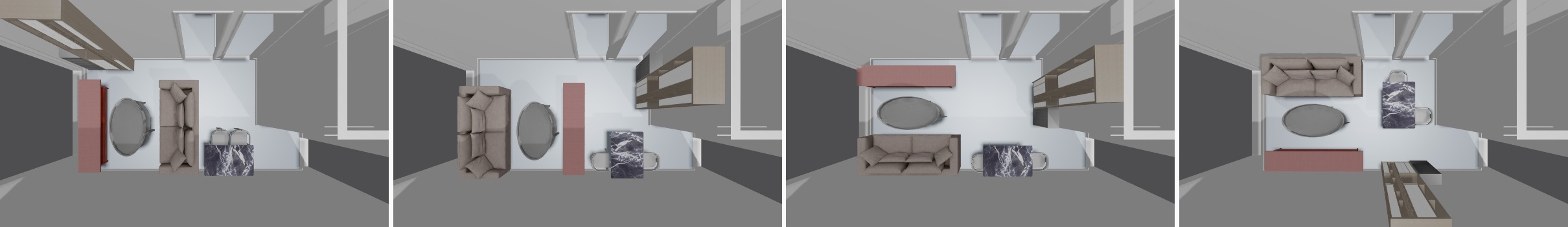}
	\end{subfigure}
	\begin{subfigure}[b]{0.49\linewidth}
		\includegraphics[width=\linewidth]{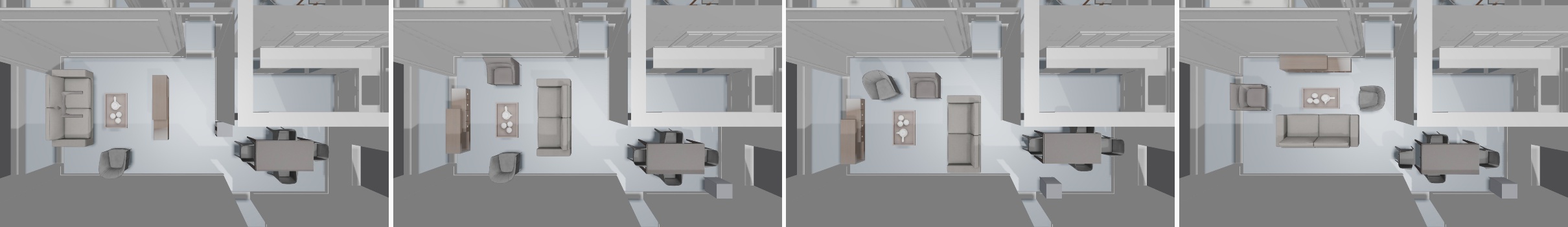}
	\end{subfigure}
	
	\begin{subfigure}[b]{0.49\linewidth}
		\includegraphics[width=\linewidth]{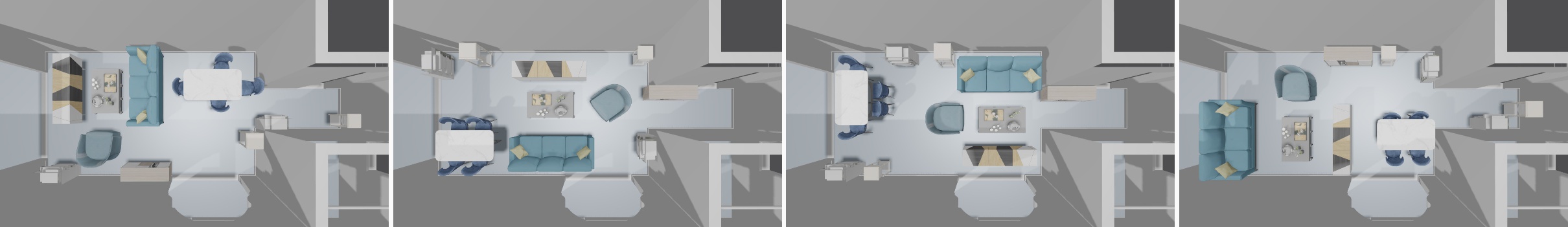}
	\end{subfigure}
	\begin{subfigure}[b]{0.49\linewidth}
		\includegraphics[width=\linewidth]{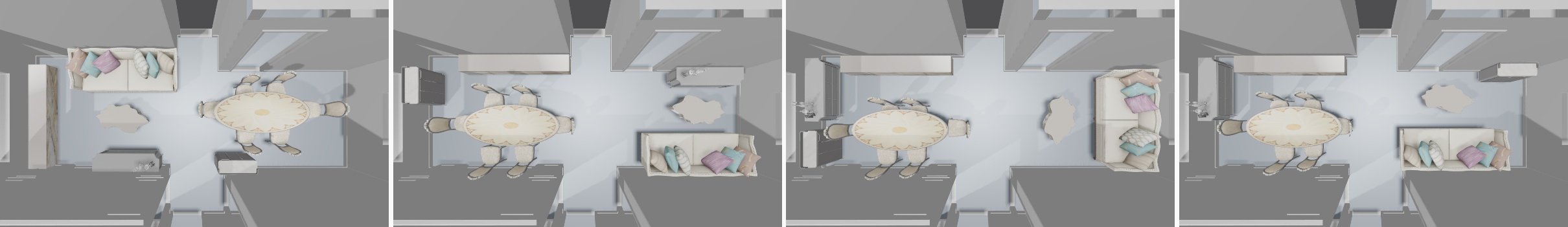}
	\end{subfigure}
	
	\begin{subfigure}[b]{0.49\linewidth}
		\includegraphics[width=\linewidth]{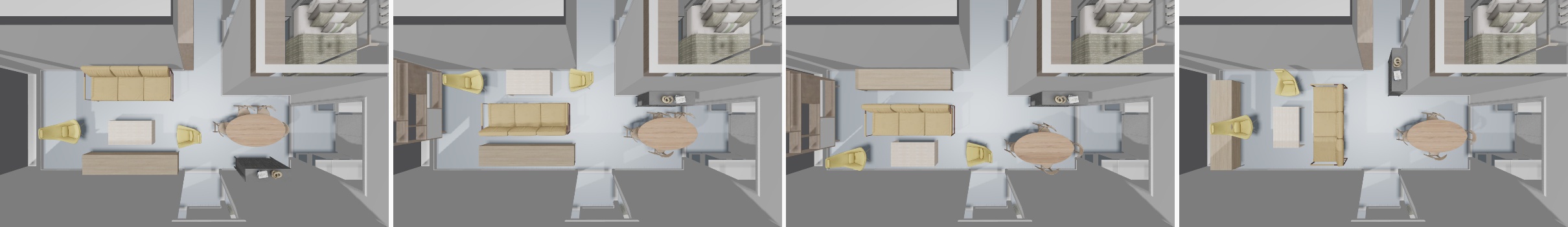}
	\end{subfigure}
	\begin{subfigure}[b]{0.49\linewidth}
		\includegraphics[width=\linewidth]{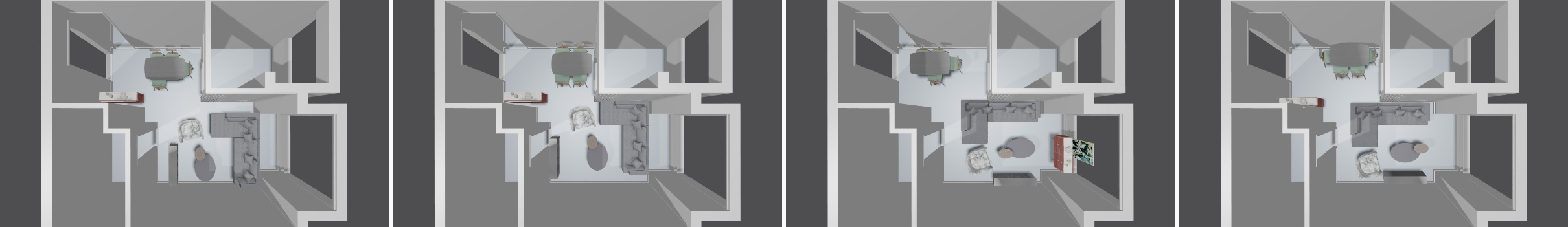}
	\end{subfigure}
	
	\begin{subfigure}[b]{0.49\linewidth}
		\includegraphics[width=\linewidth]{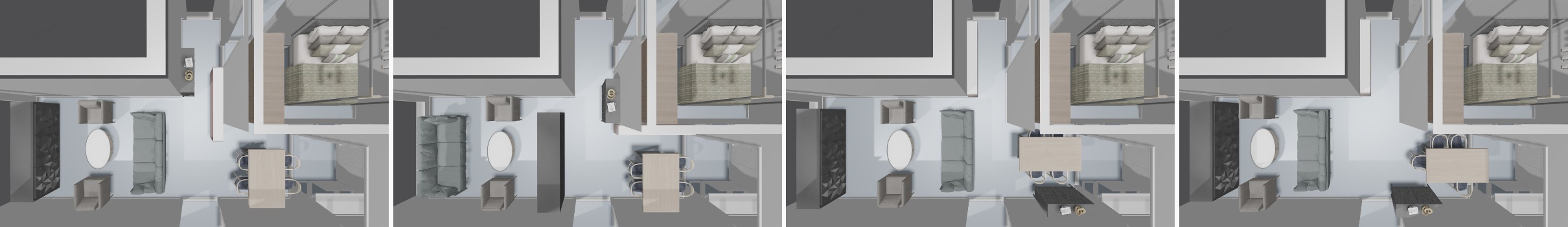}
	\end{subfigure}
	\begin{subfigure}[b]{0.49\linewidth}
		\includegraphics[width=\linewidth]{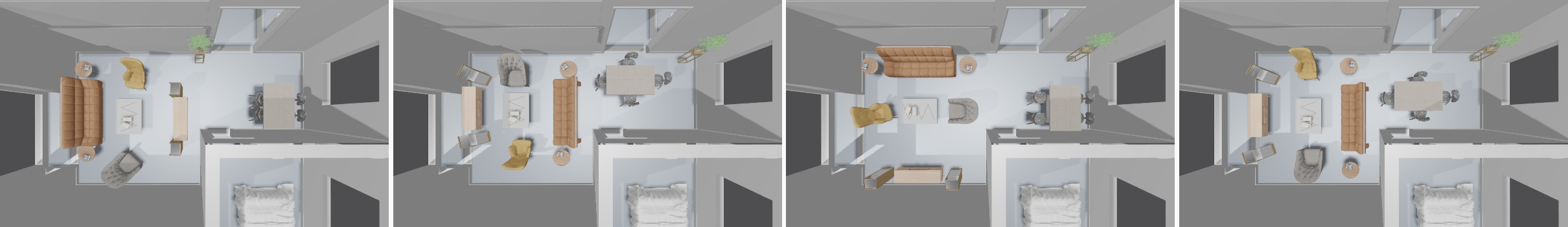}
	\end{subfigure}
	
	\vspace*{-3mm}
	\caption*{Living Room \& Dinning Room. }
	\vspace*{-3mm}
	\caption{Results. Please zoom in for more details. More results are included in the supplementary files. }
	\label{fig:results}
\end{figure*}

We utilize a recent 3D scene dataset ``3D-Front\footnote{https://tianchi.aliyun.com/dataset/dataDetail?dataId=65347}" \cite{fu20203dfuture} with 70000+ layouts and 9992 3D models. To roam and render 3D scenes, we develop an open-source 3D scene platform as shown in figure \ref{fig:platform}, where we can add, delete, modify and search objects. We can orbitally control the perspective camera for selecting better views. By clicking ``layout", configurations of a current room is layouted by our proposed framework. We render 3D scenes using Three.js\footnote{http://threejs.org/} and the algorithm is mainly implemented by PyTorch and NumPy. Several results are shown in figure \ref{fig:results}. Please refer to our supplementary materials for more details.\footnote{We also run our framework on SUNCG \cite{song2017semantic} before this dataset became unavailable. We include results of SUNCG optionally in our supplementary materials only to verify the effectiveness of our framework. }

\begin{figure*}
    \centering
    \begin{subfigure}[b]{0.325\linewidth}
		\includegraphics[width=\linewidth]{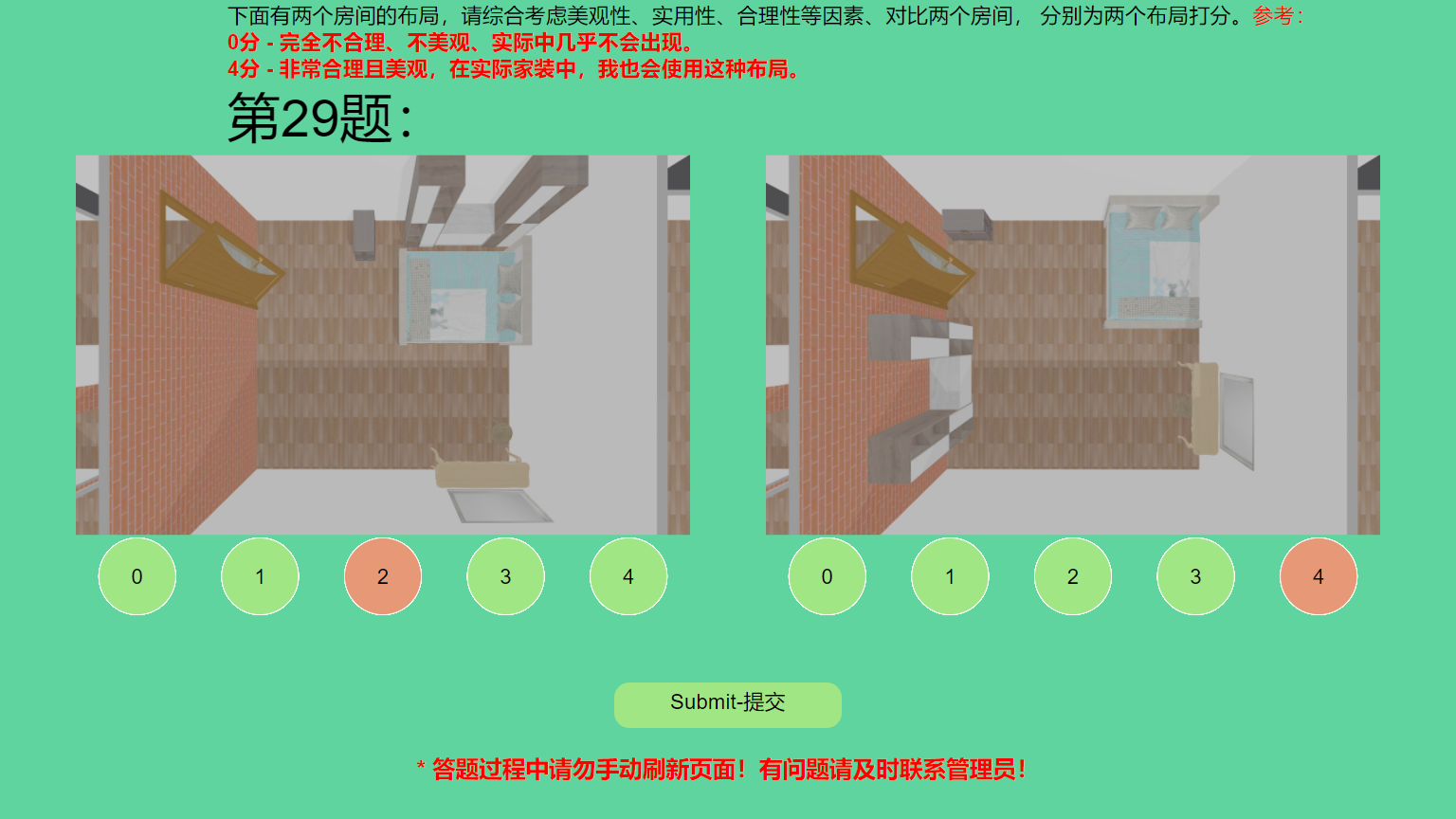}
		\caption{}
		\label{fig:compareplanit}
	\end{subfigure}
    \begin{subfigure}[b]{0.325\linewidth}
		\includegraphics[width=\linewidth]{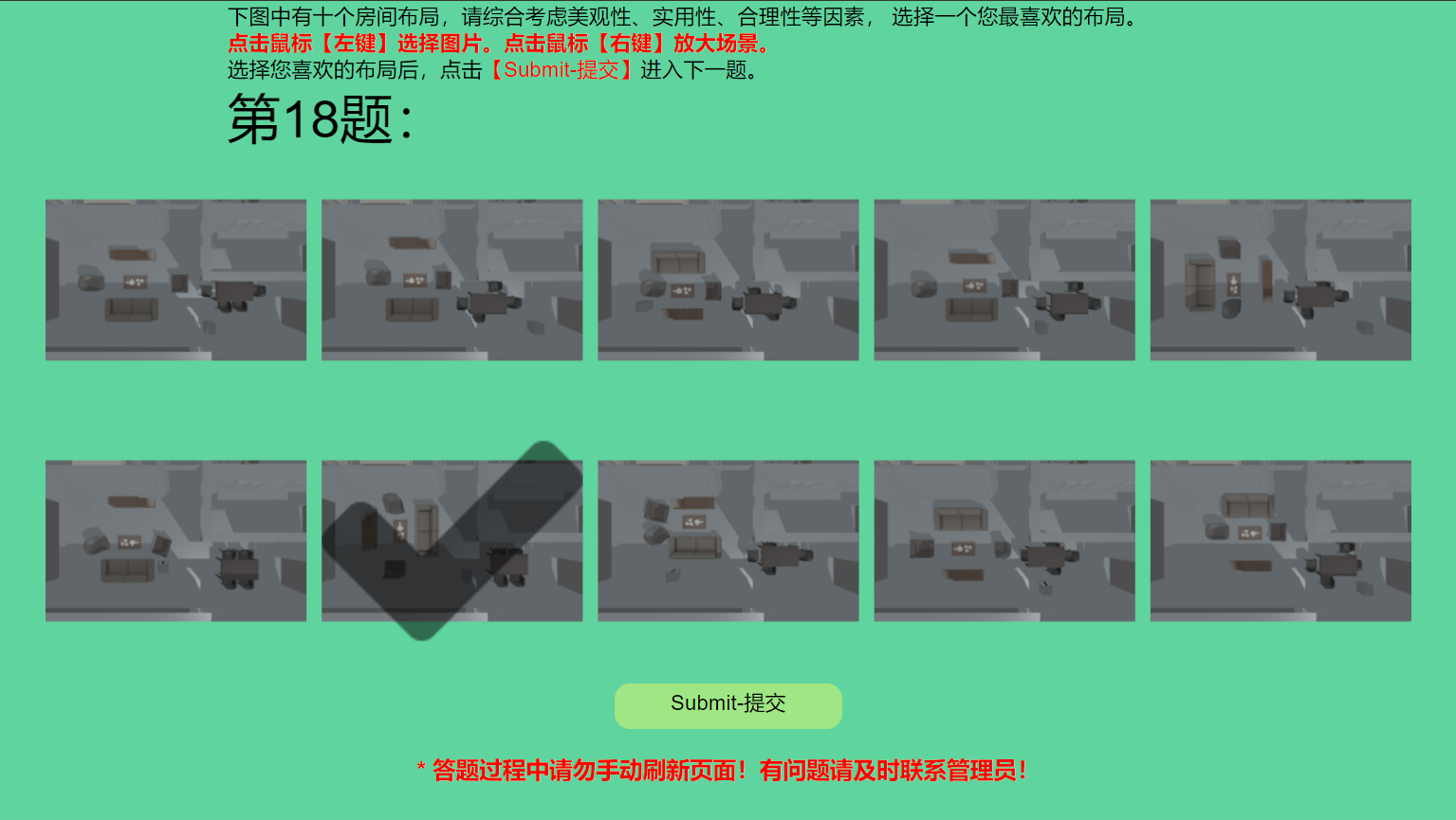}
		\caption{}
		\label{fig:select10}
	\end{subfigure}
	\begin{subfigure}[b]{0.325\linewidth}
		\includegraphics[width=\linewidth]{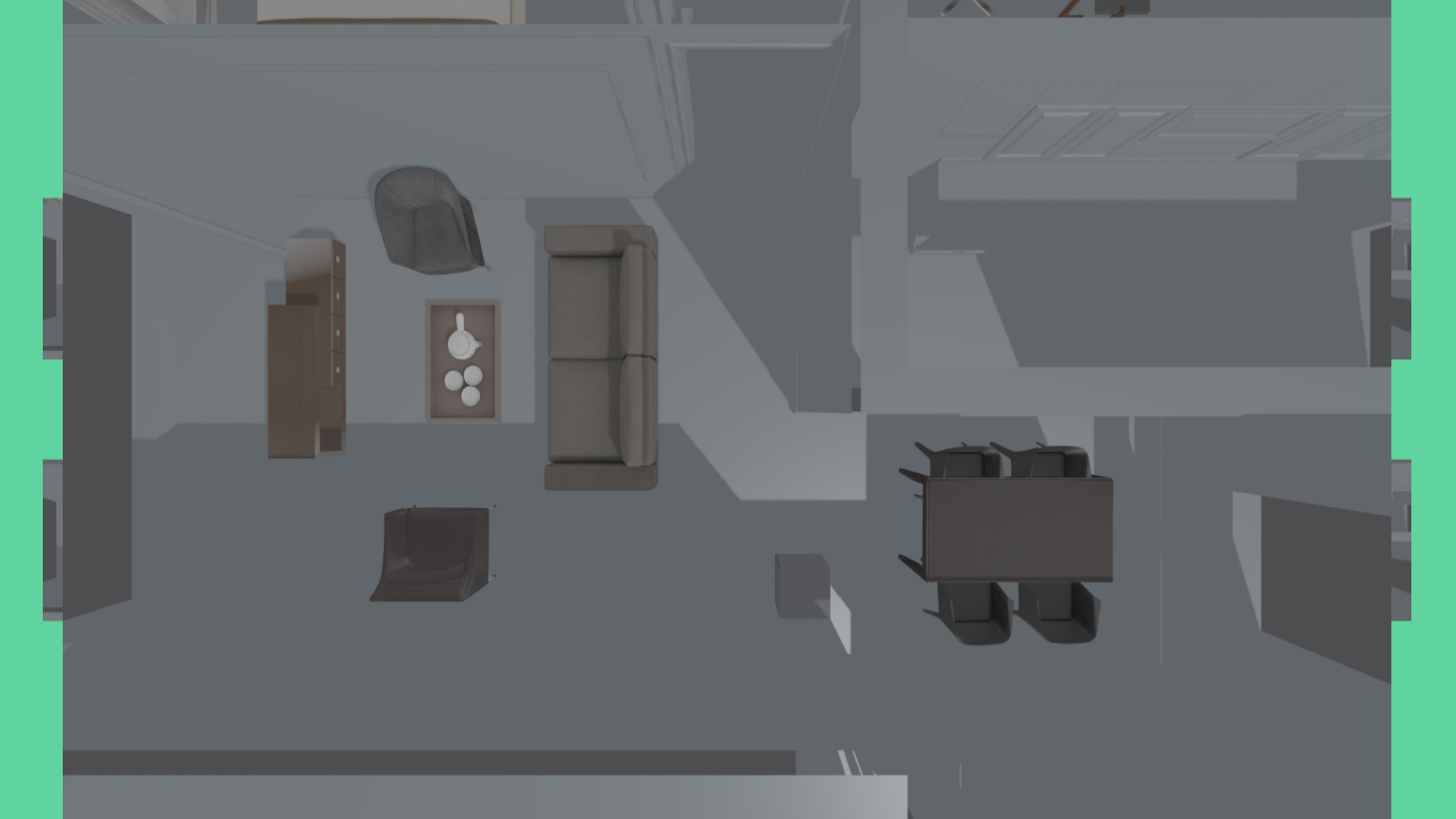}
		\caption{}
		\label{fig:zoomin}
	\end{subfigure}
    \caption{User studies. \ref{fig:compareplanit}: Marking ours and PlanIT \cite{wang2019planit} respectively; \ref{fig:select10}: Selecting the most plausible layout from ten alternative scenes where one scene is generated by human designers; \ref{fig:zoomin}: subjects can zoom in a particular layouts for better cognition. }
    \label{fig:userstudy}
\end{figure*}

\subsection{Plausibility and Aesthetic} \label{sec:compareplanit}
We compare our framework with the state-of-art PlanIT \cite{wang2019planit}. PlanIT includes not only arranging objects but also selecting appropriate objects. However, since we focus on arranging objects, we show better plausibility and aesthetic achieved using our framework by re-arranging results of PlanIT, i.e, we generate layouts given objects and room shape selected by PlanIT. 

\begin{figure}
    \centering
    \begin{subfigure}[b]{0.329\linewidth}
		\includegraphics[width=\linewidth]{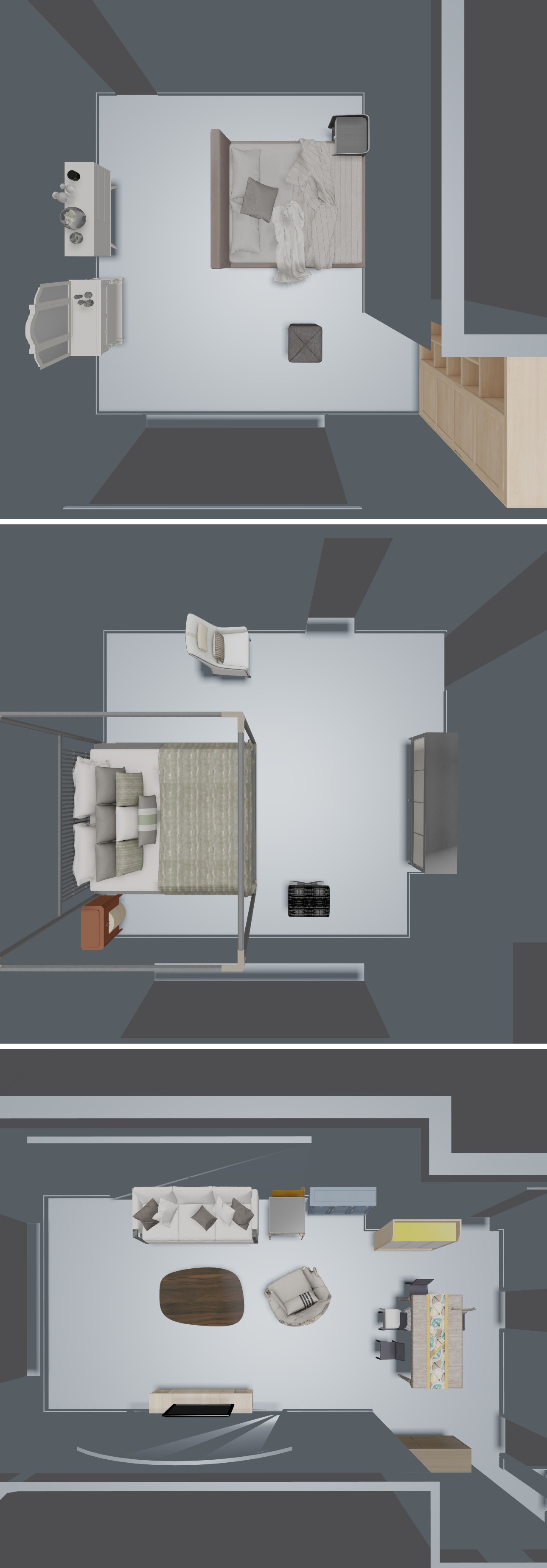}
		\caption{PlanIT. }
	\end{subfigure}
	\begin{subfigure}[b]{0.66\linewidth}
		\includegraphics[width=\linewidth]{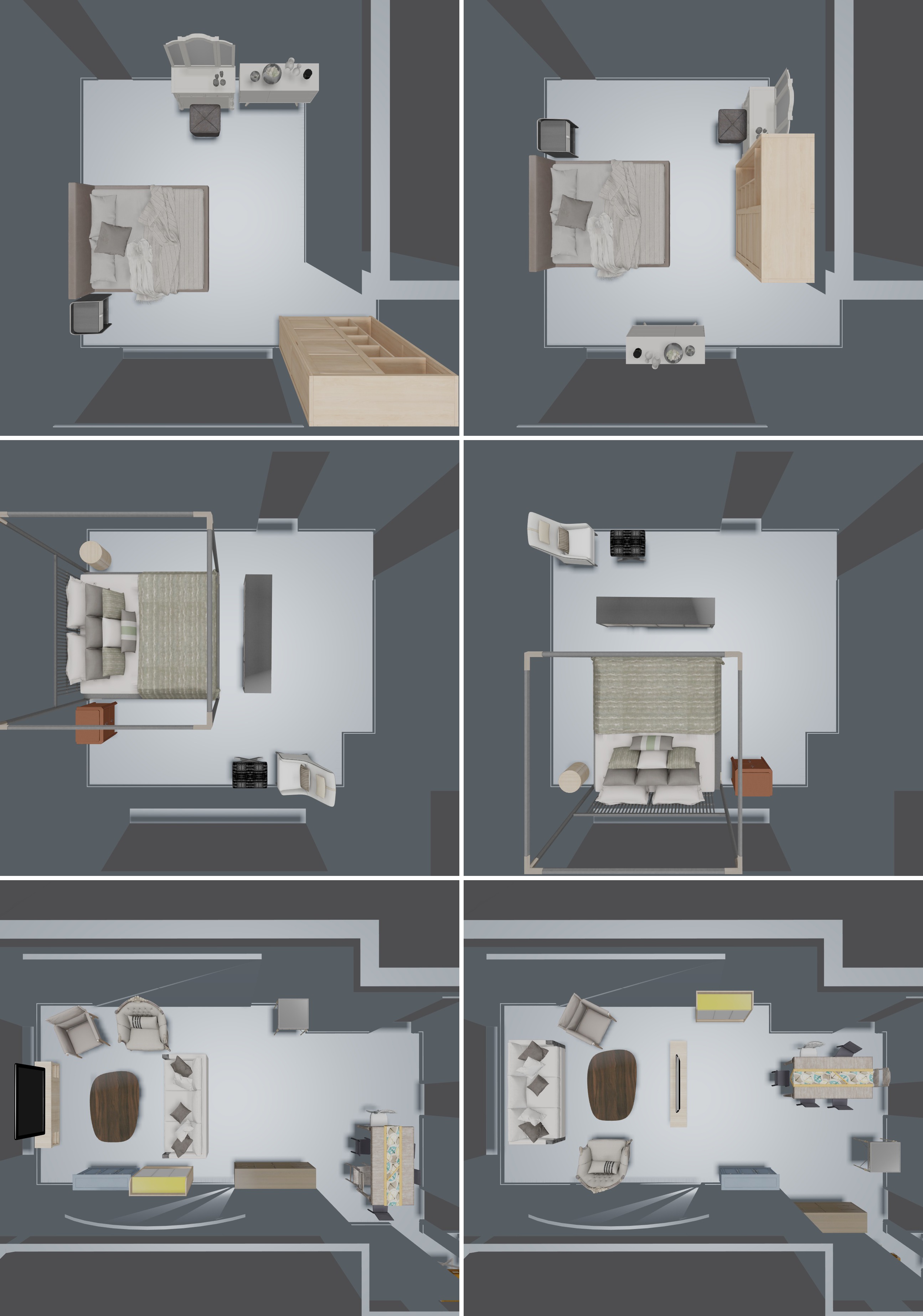}
		\caption{Ours. }
	\end{subfigure}
    \caption{Qualitatively comparing PlanIT with ours. }
    \label{fig:withplanit}
\end{figure}

Qualitatively, as shown in figure \ref{fig:withplanit}, ours is friendly for layouts among objects with strong relations, i.e., ``coherent groups" in this paper. For example, a TV stand and a sofa are strongly related to a coffee table. Ours makes sure they are plausibly arranged among each other. Additionally, ours does not block paths of doors and windows. Quantitatively, we also conduct a user study as shown in figure \ref{fig:compareplanit}. 43 subjects are invited. Subjects are university students, workers, housewives, interior designers, etc.\footnote{Few subjects preserve privacy. } Each subject is given 20 questions and each question includes a layout generated by ours and a layout generated by PlanIT. Presented layouts are shuffled, i.e., ours can be either left or right. For each question, a subject compares two layouts and marks them respectively. Marks ranged from $0$ (very poor) to $4$ (very plausible). All subjects are taught how to use the user study system before experiencing. \sk{In figure \ref{fig:compareplanit}, the Chinese characters are rendered as ``there are two room layouts below, please compare the two layouts, considering aesthetic, plausibility and reasonableness, thus marking them respectively.", ``0: totally unreasonable, inaesthetic. It may never appear in the real world layout. " and ``5: very aesthetic and plausible. I will refer to this layout in the real world."} Results are shown in table \ref{tab:compareplanit}, where marks are averaged (standard deviation) of respective room types. 

\begin{table}
    \caption{User study: results of comparing PlanIT with ours. }
    \label{tab:compareplanit}
	\begin{center}
		\begin{tabular}{|c|c|c|}
			\hline
			Room Type & PlanIT & Ours \\
			\hline\hline
			Bedrooms     & 1.847 (1.336)  &  2.66 (1.125) \\ 
			Living Rooms & 1.749 (1.327)  &  2.572 (1.266) \\ 
			Bathrooms    & 1.028 (1.2)  &  2.553 (1.314) \\ 
			Kitchens     & 1.549 (1.342)  &  2.651 (1.167) \\ 
			Total        & 1.543 (1.341)  &  2.609 (1.221) \\ 
			\hline
		\end{tabular}
	\end{center}
\end{table}

\subsection{Robustness}
\begin{figure}
    \centering
    \includegraphics[width=\linewidth]{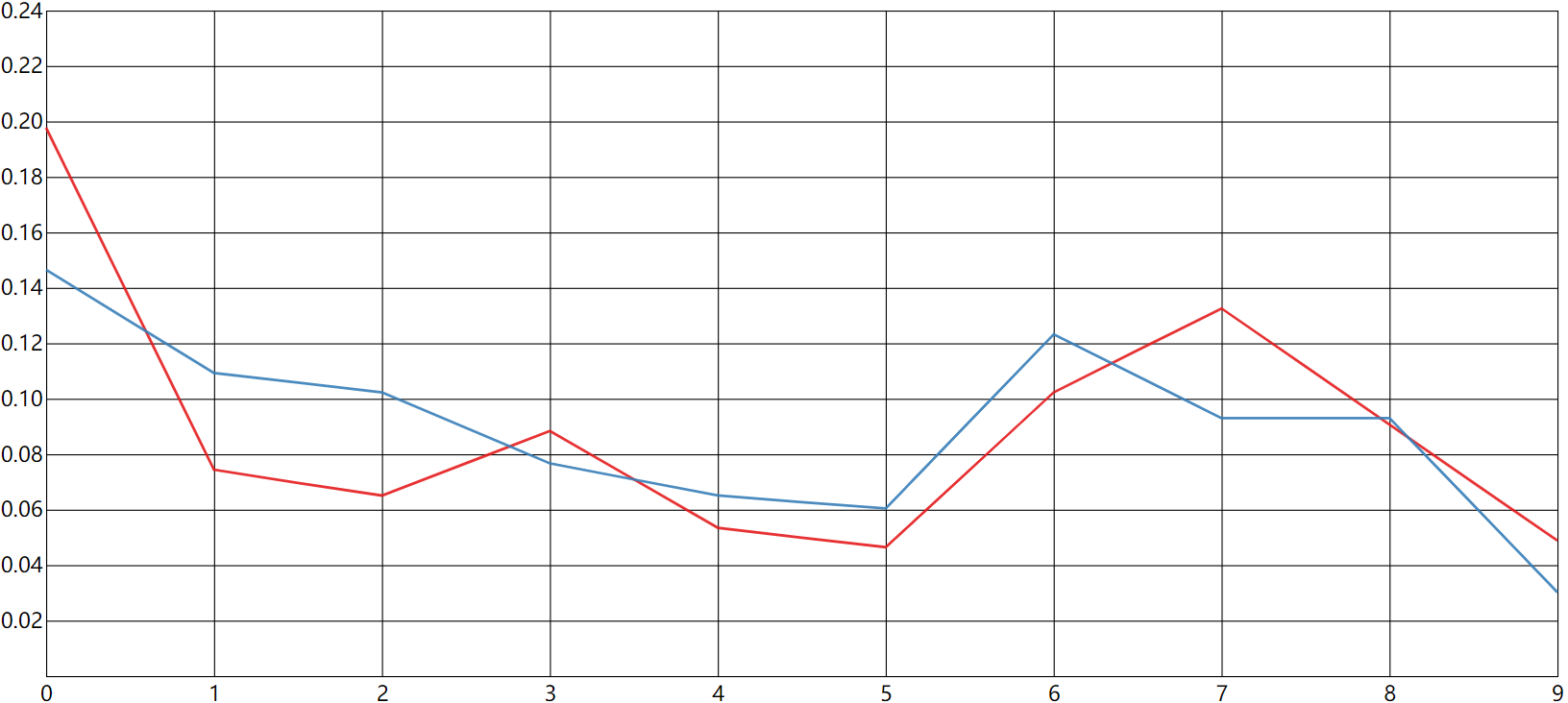}
    \caption{Distributions of user selected layouts of bedrooms (BLUE) and ``living-dinning rooms" (RED). }
    \label{fig:robustness}
\end{figure}

In this section, we compare our generated layouts with the layouts of human designers (ground truths) to verify that ours is competitive to human. Subjects are the same group of section \ref{sec:compareplanit}. Each subject is required to choose a most plausible layout from ten alternative layouts as shown in figure \ref{fig:select10}, where one layout is designed by a human designer and remaining nine layouts are generated by ours. Subjects can zoom in layouts by right clicks such as figure \ref{fig:zoomin}. All subjects are taught before experiencing and manuals are available. Ground truths are randomly selected from 3D-Front. \sk{In figure \ref{fig:select10}, the Chinese characters are rendered as ``there are ten layouts below and please select your favorite one considering aesthetic, plausibility and reasonableness", ``left-click for selections and right click for zooming in" and ``after selecting, press submit for the next question".} 

Results are shown in figure \ref{fig:robustness}. Two distributions are plotted for bedrooms and ``living-dinning" rooms respectively, i.e., each line is averaged distributions of user selections of its room type, where ``0" denotes ground truth. Although human-designed layouts outperform ours, generated layouts are still favored competitively as shown in figure \ref{fig:robustness}. 

\subsection{Efficiency}

We experience our framework on a PC with AMD 2700X (GHz), GTX 970, and WD20EZRX. Time consumption of layouts depend on degrees of crowding, i.e., ratio of total area of coherent groups to area of room. Higher degrees result in more discards during geometry-based arrangements (section \ref{sec:wallchasing}), thus slowing down generations. 

To layout 3D-Front such as figure \ref{fig:results}, if priors are cached, our framework consume within $3.5$ second for a layout. If corresponding priors of several objects are not loaded, additional IO is required up to $2$ seconds for a layout. For non-crowded rooms, with cached priors, our framework generates layouts in real time. 

We also run the state-of-art PlanIT \cite{wang2019planit} on servers with GTX 1080ti. According to our experiments, generating a layout requires more than a minute. Nevertheless, this includes both object selection and object arrangement and the two are interleaved with each other. Testing exact time consumption of ``layout" of PlanIT is beyond the scope of this paper. Furthermore, \cite{weiss2018fast} is not a data-driven framework. Therefore, it is hard to conclude ``better efficiency" as a contribution. 

\section{Conclusions and Future Works}
In this paper, we present a new framework of generating room layouts and we experimentally verify the achieved plausibility and robustness. The code of this framework and a toolbox platform is available. We hope this could contribute to domains related to 3D scenes. However, this work still suffers the following weaknesses. 

\begin{figure}
    \centering
    \begin{subfigure}[b]{0.325\linewidth}
		\includegraphics[width=\linewidth]{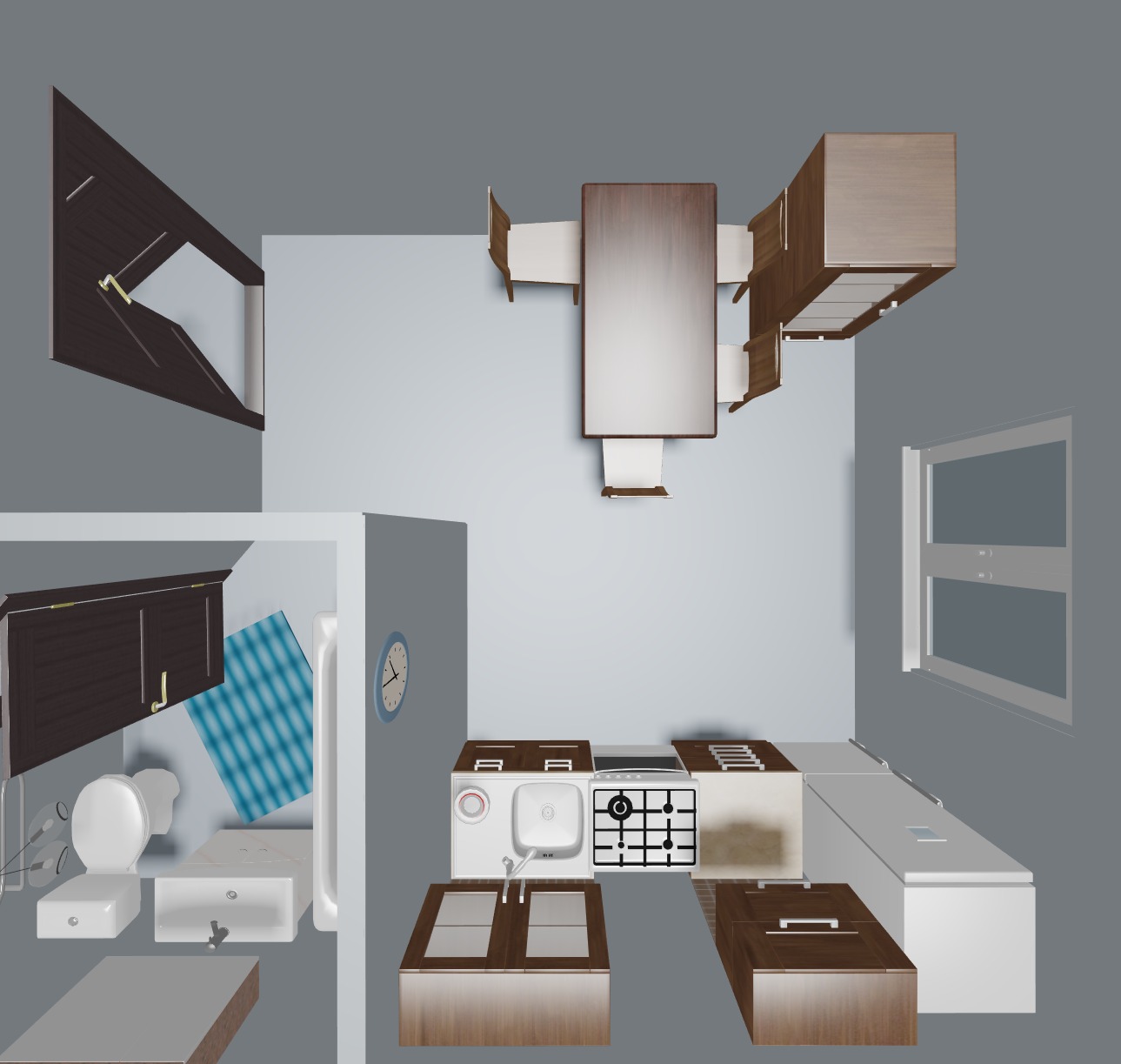}
	\end{subfigure}
	\hfill
	\begin{subfigure}[b]{0.325\linewidth}
		\includegraphics[width=\linewidth]{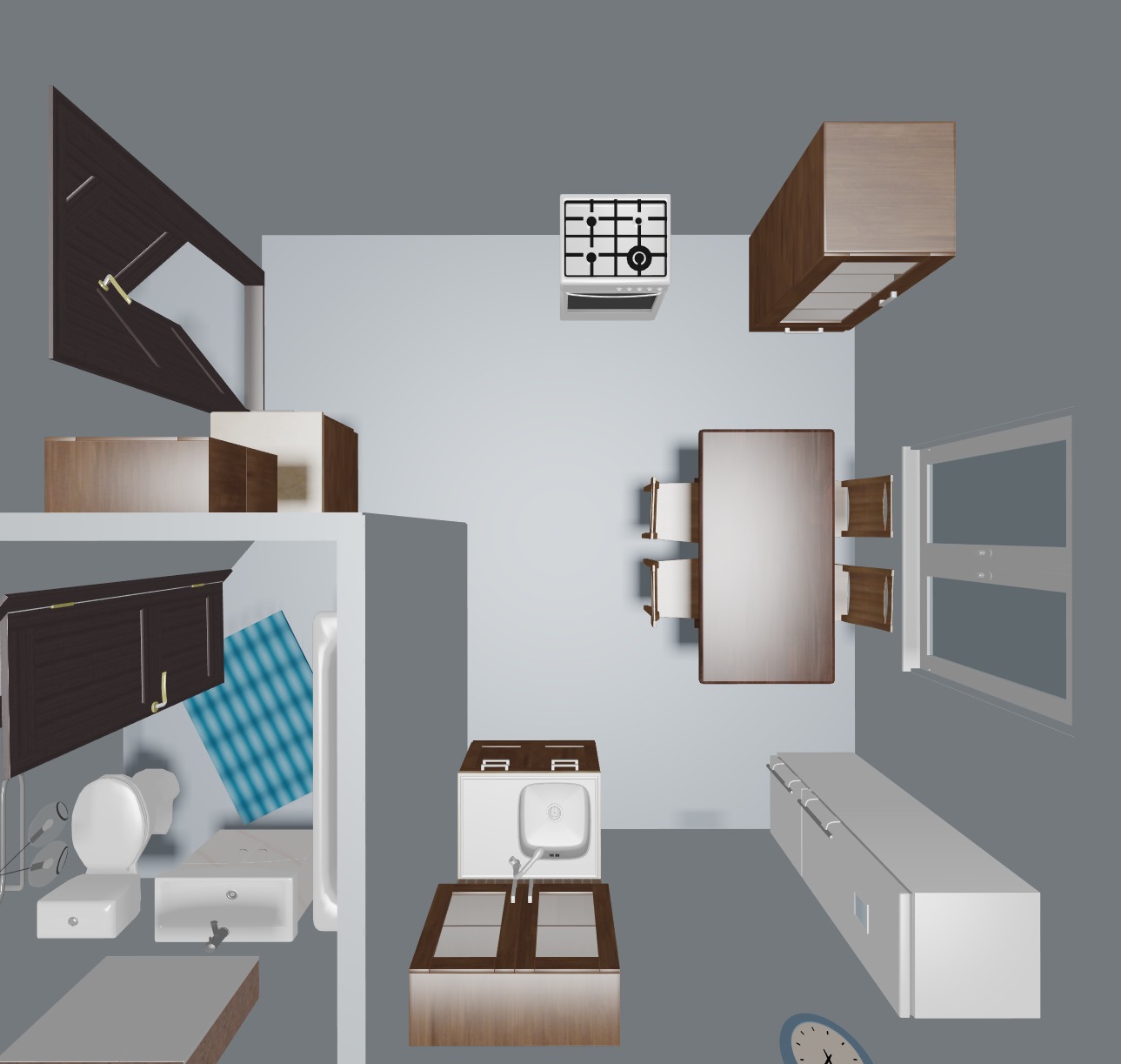}
	\end{subfigure}
	\hfill
	\begin{subfigure}[b]{0.325\linewidth}
		\includegraphics[width=\linewidth]{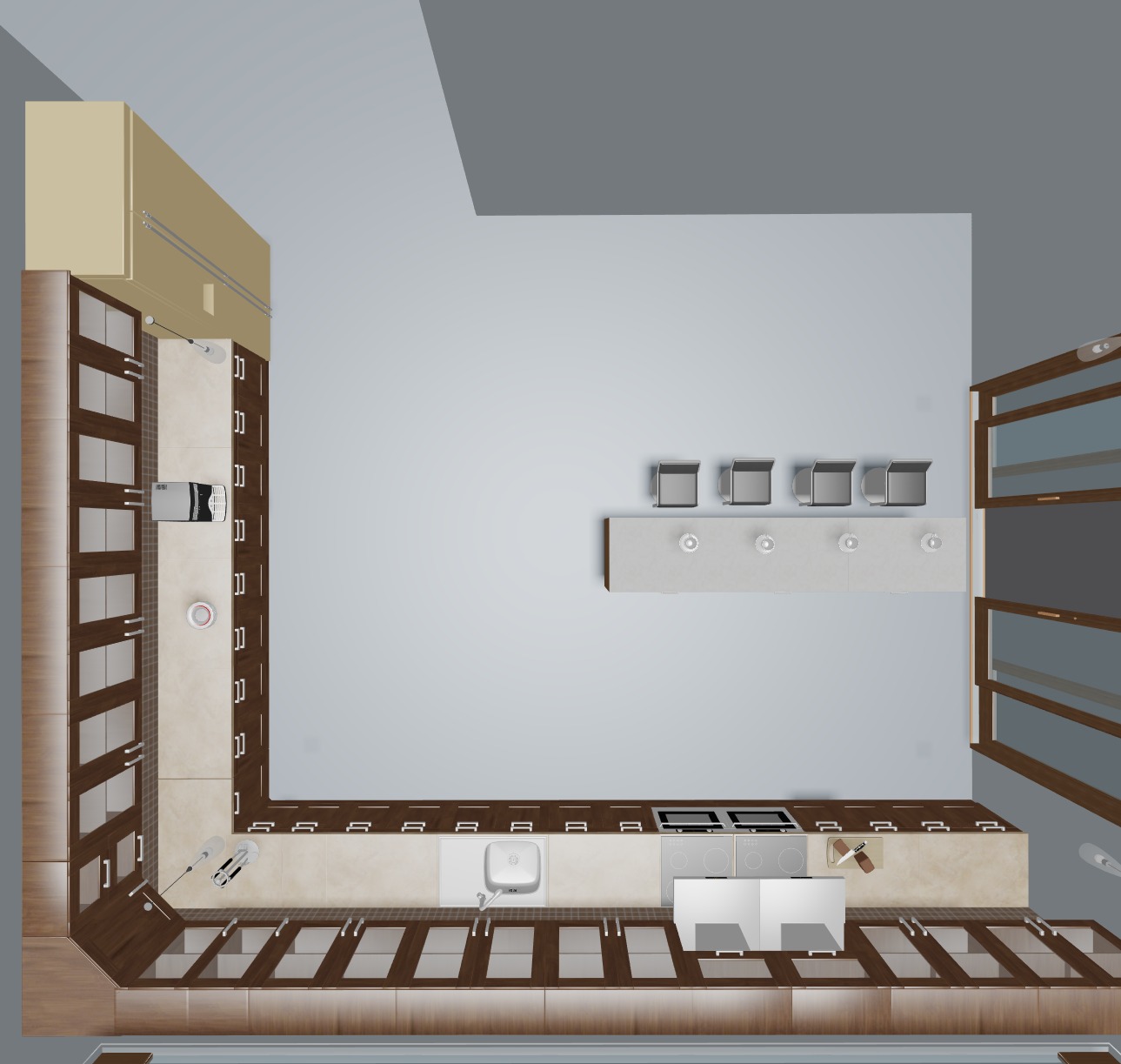}
	\end{subfigure}
    
	\begin{subfigure}[b]{0.325\linewidth}
		\includegraphics[width=\linewidth]{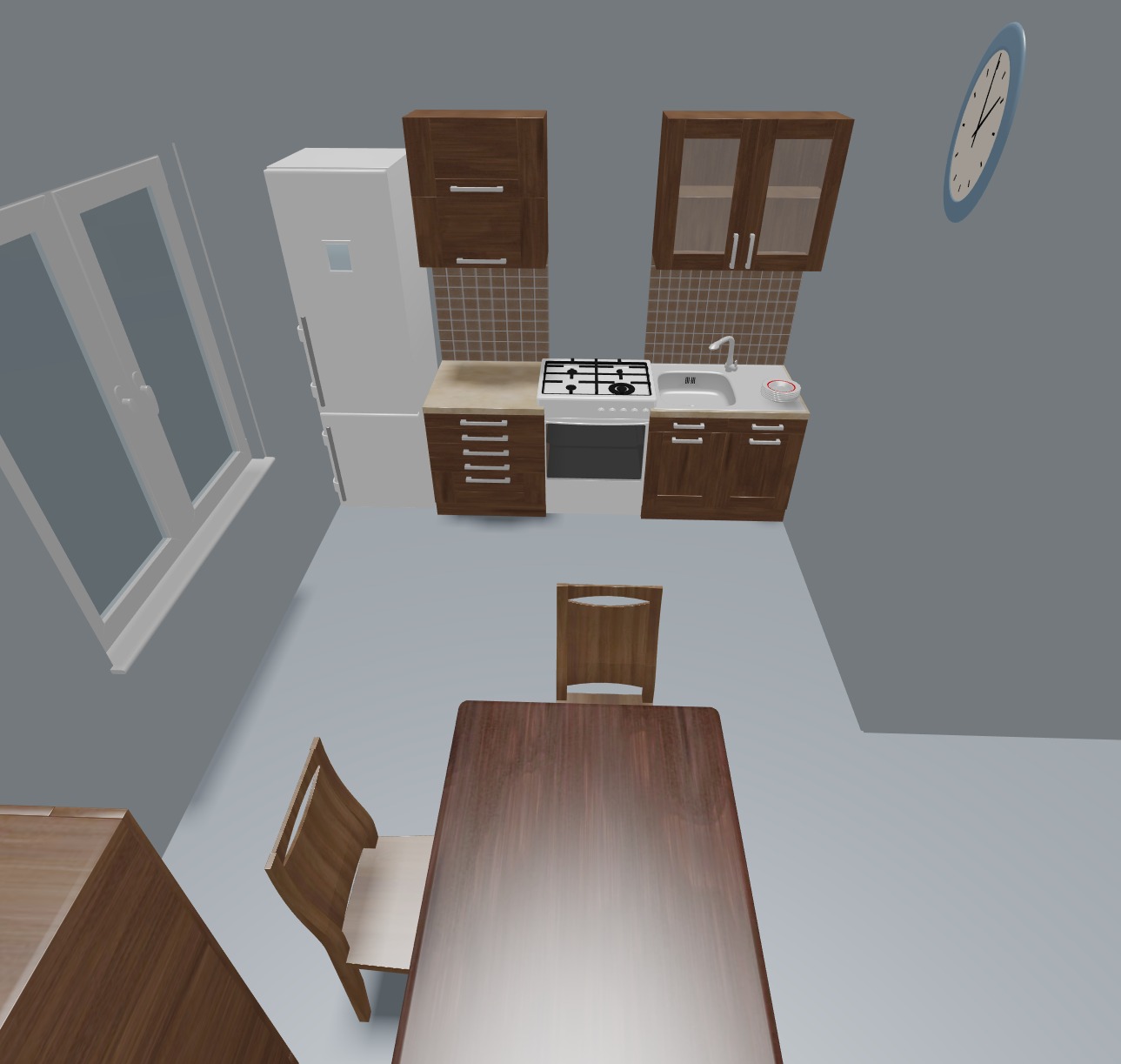}
		\caption{}
		\label{fig:conclusionchain1}
	\end{subfigure}
	\hfill
	\begin{subfigure}[b]{0.325\linewidth}
		\includegraphics[width=\linewidth]{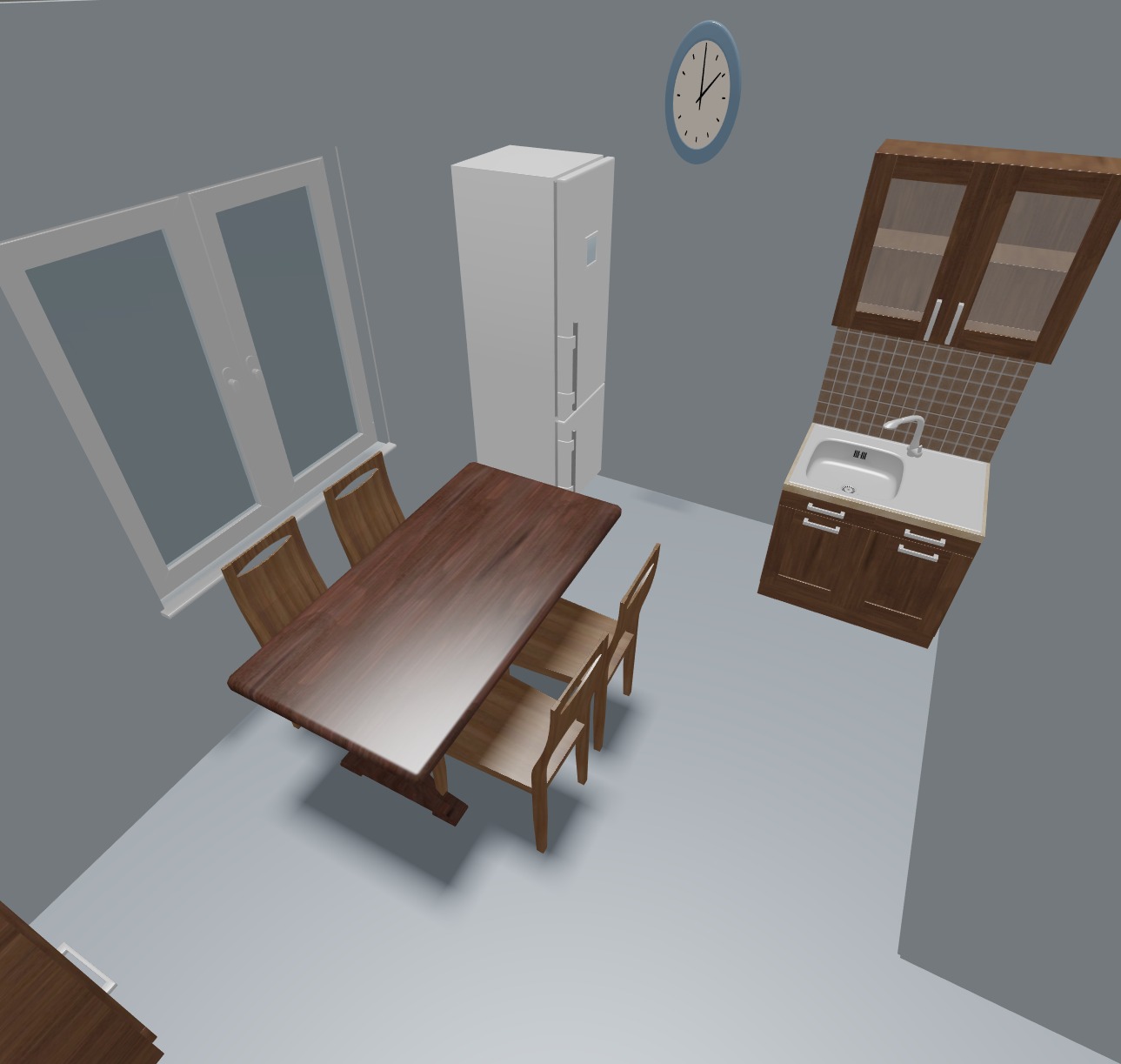}
		\caption{}
		\label{fig:conclusionchain2}
	\end{subfigure}
	\hfill
	\begin{subfigure}[b]{0.325\linewidth}
		\includegraphics[width=\linewidth]{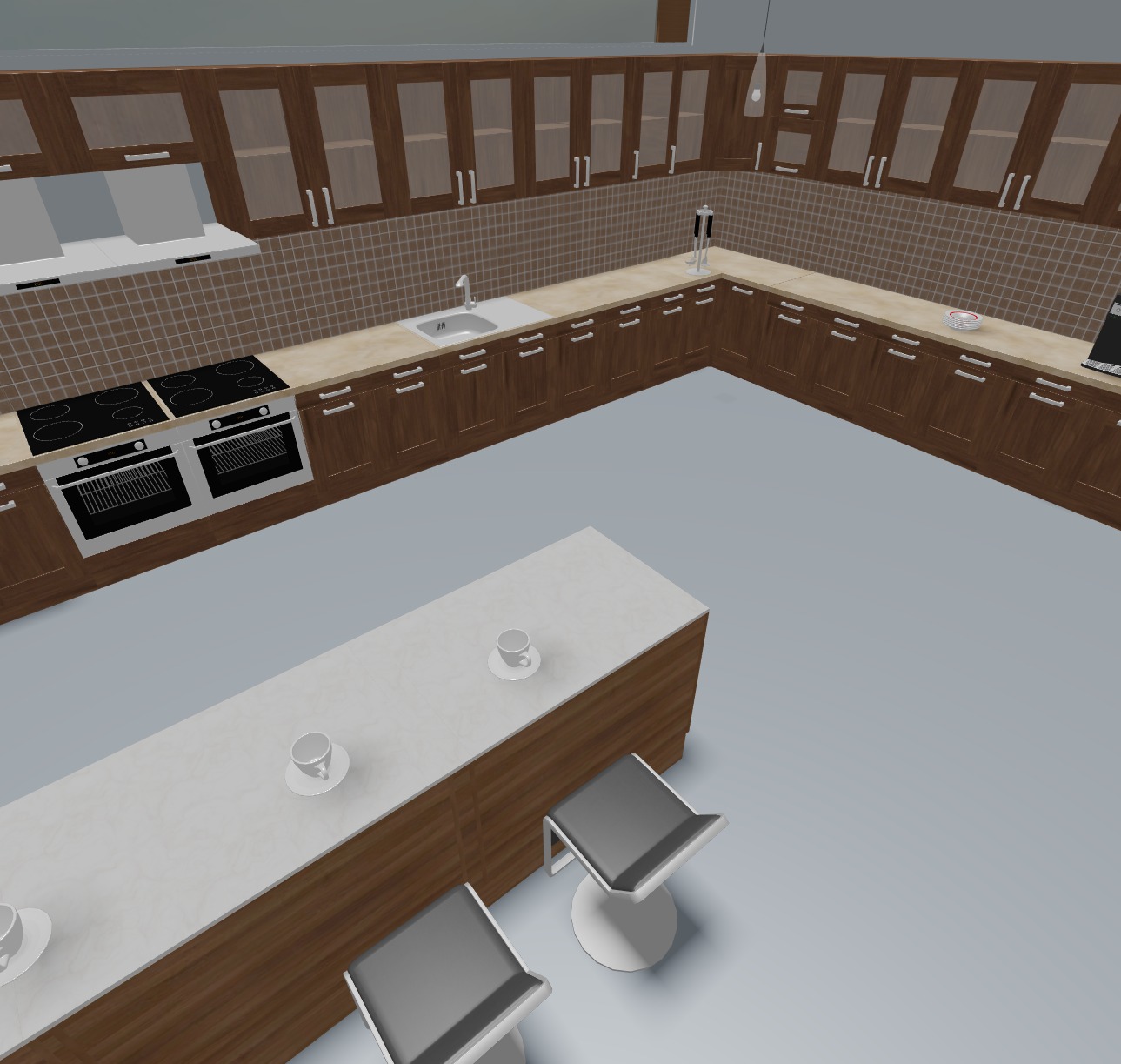}
		\caption{}
		\label{fig:conclusionchain3}
	\end{subfigure}
    \caption{The problem of ``chains" of objects around walls. \ref{fig:conclusionchain1}: The ground truth; \ref{fig:conclusionchain2}: A failure case of ours; \ref{fig:conclusionchain3}: An L-shape ``chain" with L-shape objects. }
    \label{fig:conclusionchain}
\end{figure}

The most difficulties we encountered is arranging ``chains" of objects around walls. For independent objects such as wardrobes, transformations of them have high degree of freedom since we find appropriate places for them with no collision and implausibility. However, for groups of objects such as kitchen cabinets and ovens, they are frequently placed adjacently next to each other as shown in figure \ref{fig:conclusionchain}. Firstly, orders of a chain should be carefully considered. For example, commonly we places geometrically similar cabinet next to each other. Otherwise, layouts are not aesthetic as shown in figure \ref{fig:conclusionchain2}. Secondly, an L-shape chain should somehow turn at corners, especially when we have L-shape objects such as L-shape cabinets which are frequently treated as ``corner objects" as shown in figure \ref{fig:conclusionchain3}. Thirdly, doors and windows are also challenges for arranging chains. In our framework, if we treat a chain as an entire group, currently we do not have plans for sampling such priors. On the other hand, if we treat a chain as individual objects, complicated rules are required but we also do not have a plan for formulating the rules. As a result, we demonstrate this weakness in detail and we would try fixing it in future. Fortunately, in real-world decoration, most cabinets are fixed on walls. 

The storage and loading of priors may require further system-level optimizations. Currently, all priors are structured in ``.json" format, which is inefficient if a prior of a coherent group is too large. When arranging objects online, loading priors may consume up to few seconds for loading corresponding priors into the memory. Although this only affects the first attempt, since priors are cached after that, it is still a concern in practice. Eventually, the way of extracting patterns is robust to shapes and textures instead of incorporating them. This is useful if we want objects arranged more compacted. 

{\small
\bibliographystyle{cvm}
\bibliography{cvmbib}
}

\end{document}